\documentclass{ws-procs9x6}
\usepackage{epsfig}
\begin{document}

\title{TASI Lectures on Neutrino Physics}

\author{ANDR\'E de GOUV\^EA}

\address{Northwestern University, Department of Physics \& Astronomy \\
2145 Sheridan Road, \\ 
Evanston, IL 60208, USA\\ 
e-mail: degouvea@northwestern.edu}

\maketitle

\abstracts{
I discuss, in a semi-pedagogical way, our current understanding of neutrino physics. I present a brief history of how the neutrino came to be ``invented'' and observed, and discuss the evidence that led to the recent discovery that neutrinos change flavor. I then spend some time presenting mass-induced neutrino flavor change (neutrino oscillation), and how it pieces all the neutrino puzzles except for the LSND anomaly, which is also briefly discussed. I conclude by highlighting the importance of determining the nature of the neutrinos, {\it i.e.}, are they Dirac or Majorana fermions.}

\hfill  NUHEP-TH/04-17

\section{Introduction}

Neutrino physics is among the most exciting and active areas of research in high energy physics, both experimentally and theoretically. No other area of particle physics research has gone through as dramatic a transformation, and one can arguably claim that the near and intermediate future of research associated with neutrinos is very bright indeed. 

The reason for all the excitement is the fact that a plethora of neutrino detectors have obtained, without a shadow of a doubt, the only palpable evidence for physics beyond the standard model of eletroweak interactions. New physics seems to have manifested itself in the form of neutrino masses and lepton mixing.  

Here, I discuss not only what we know (and don't know) about neutrino properties and interaction, but also spend some time exploring how these were uncovered experimentally and theoretically. In Sec.~\ref{history}, I present a condensed history of the neutrino, including how they were discovered, first theoretically, and then experimentally, and when we started suspecting that there was something wrong with the naive standard model description of neutrino properties. In Sec.~\ref{oscillations}, I discuss in some detail the phenomenon of neutrino oscillations, and describe how oscillations were used to solve almost all long-standing neutrino puzzles in Sec.~\ref{solve_puzzle}. Sec.~\ref{sec_lsnd} contains a brief digression on the still-unresolved LSND anomaly. Finally, in Sec.~\ref{theend}, I summarize what we have learned from neutrinos, what I mean by ``neutrinos provide the only palpable evidence of physics beyond the standard model,'' and what we think this new physics is.

Before proceeding, two comments. (i) In response to the rapid metamorphosis of neutrino physics, several excellent overviews and pedagogical lecture notes have been written and made available to the community,\cite{reviews,review_history,book} and these discuss at different levels of detail and rigor everything contained here. I hope that this ``historical approach'' to the subject adds to the literature, which is quite excellent. (ii) I apologize in advance for all inaccuracies regarding historical facts and the omission of several important references.      

\section{Beta-Rays and Missing Neutrinos}
\label{history}

Here I review, in a very biased, simplified, and condensed form, the history of weak interactions and neutrino physics. Not only is it a fascinating subject, but it also serves to illustrate that neutrinos have always managed to challenge and confuse physicists into contemplating what was once perceived to be impossible and questioning fundamental principles and theoretical biases. Over the years, neutrinos have provided fundamental clues that lead to significant progress in the understanding of fundamental physics. I also wish to make the point that progress in neutrino research had always been very slow, in contrast to the boom observed in the past six years. I refer readers to Ref.~\refcite{book} for a detailed overview of the history of neutrinos. 

I also summarize the series of experimental results that lead to the discovery that neutrinos can change flavor, hence changing in qualitative way our understanding of them.

\subsection{Biased Neutrino History}

The history of the neutrino\footnote{Most of the material discussed in this subsection was obtained from Ref.~\refcite{book}.} is strongly tied to the history of the once very mysterious weak interactions. These were ``stumbled upon,'' quite accidently, by Henri Becquerel in 1896, one year before the electron was discoverd by J.J.~Thompson and several years before nuclei were known to exist. All pioneering studies of natural radioactivity also took place years before the development of even the most rudimentary aspects of quantum mechanics. 

Becquerel is credited with the discovery that, even in the absence of an outside stimulus (e.g. light, electricity) some chemical elements naturally emit radiation capable, for example, of leaving a mark in a photographic plate. In the early 1900's, detailed analysis of this natural radioactivity revealed that different elements emitted different types of radiation, and these were baptized, by Lord Rutherford, $\alpha$, $\beta$, and $\gamma$ radiation, or rays. The different types of radiation possessed rather distinct properties, and it is curious to note, in hindsight, that each was due to a different interaction ($\alpha\to$~strong, $\beta\to$~weak, $\gamma\to$~electromagnetic):
\begin{itemize}
\item $\alpha$-rays were easy to absorb and bent slightly (positive charge, ``heavy'') in the presence of magnetic fields. It was discovered, due to their absorption properties, that $\alpha$-rays emitted from a well defined isotope were mono-energetic;
\item $\beta$-rays were harder to absorb than $\alpha$-rays, and they bent significantly (negative charge, ``light'') in the presence of magnetic fields;
\item $\gamma$-rays did not bent in the presence of magnetic fields (no charge), and were very hard to absorb.
\end{itemize}

\subsubsection{The Nuclear $\beta$-Decay Spectrum}

The pre-history of the neutrino begins with detailed studies of $\beta$ radiation. Progress in the understanding of $\beta$-rays was very slow and rather ``tortuous.'' 

Early studies revealed that $\beta$-rays were identical to cathode rays (electrons), and that their spectrum was discrete, like the spectrum of $\alpha$-radiation. While the former is correct, the later incorrect understanding was obtained from two distinct pieces of evidence: (i) studies of the absorption of $\beta$-rays seemed to indicate that these were mono-energetic, and (ii) the spectrum of a $\beta$-ray beam in the presence of a constant magnetic field incident on a photographic plate seemed to be discrete, and hence indicative that the $\beta$-beam was composed of several discrete components, each with a different, well-defined energy.    
 
Evidence (i) was shown to be incorrect for ``theoretical reasons.'' It turned out that the phenomenology used to describe $\beta$-ray absorption was wrong. It was established that, contrary to initial prejudice, electrons and $\alpha$-particles do not lose energy while propagating inside matter in the same way. Evidence (ii) was also incorrect, this time for ``experimental reasons.'' It turned out that photographic plates were not the best detector technology to observe the $\beta$-ray spectrum! 

In 1914, Chadwick, using more advanced detection techniques,  presented definitive evidence that the {\sl observed} $\beta$-ray spectrum was continuous. It remained to determine whether the {\sl primordial} spectrum was also continuous or whether it was originally discrete but somehow modified in its way from the material to the detector. Several hypothetical energy-loss mechanisms were raised, including interactions of the $\beta$-radiation with atomic electrons, or interactions with other types of radiation, including $\gamma$-rays.\footnote{Most $\beta$-ray emitters also emit $\gamma$-rays.} It was only fifteen years after Chadwick's discovery that the issue was experimentally settled: the nuclear $\beta$-ray spectrum is continuous. And no-one had any idea why that would be the case.

It is important to appreciate that it took over thirty years to establish that the $\beta$-ray spectrum was continuos, and that the theoretical prejudice of the time was that the spectrum should be discrete. The reasons behind the incorrect preconceived idea are easy to understand. First, $\beta$-radiation was a quantum mechanical phenomenon, and  quantum mechanical spectra were discrete (e.g. atomic spectra). Second of all, the $\beta$-radiation phenomenon was described by $^AZ\to ^A\hspace{-1mm}(Z+1)+e^{-}$, where $^AZ$ is a nucleus with atomic number $Z$ and mass number $A$. Energy and momentum conservation dictate that, modulo finite temperature effects (small, calculable), the electrons leave the nucleus with a fixed, well-defined energy, given by the mass difference between the parent and daughter nuclei.  It seemed, at the time, that, in order to understand continuous nuclear $\beta$-decay spectra, one was required to give up the principle of conservation of energy!

\subsubsection{1920's Nuclear Physics}

Of course, the early third of the twentieth century was not without its share of other fundamental physics puzzles. One was associated with understanding the nucleus and its constituents, and apparent contradictions of the spin-statistics theorem. In early nuclear physics, it had been postulated that nuclei were made up of protons and electrons, such that $^AZ$ contained $A$ protons and $A-Z$ electrons (such that the nuclear charge was $Z$).\footnote{There were more complicated models for the nucleus at the time which included, for example, $\alpha$-particles as fundamental nuclear constituents.} For example, an $^4$He nucleus was to be composed of four protons and two electrons, while $^{14}$N$=14p+7e^{-}$. 

There were several fundamental problems with this model for the nucleus. One was related to the magnetic moment of nuclei. It was well known at the time that the electron magnetic moment was much larger than the proton one, and also much larger then the nuclear one. How was that possible, if the nucleus was a collection of protons and electrons? Worse, perhaps, was the fact that the spin-statistics theorem seemed to be violated: according to the proton+electron nuclear model, $^{14}$N is predicted to be a collection of an odd number of fermions (14+7=21), which would mean it should have a half-integer spin and hence behave like a fermion. Experiments had revealed, however, that $^{14}$N was a boson.

Short of assuming that the above model for the nucleus was simply wrong (which turned out to be the case), there were only a few possible solutions to the puzzles raised above. One ``common theme'' to all the puzzles was that nuclear-bound electrons did not behave like unbound ones. At around 1929--1930, Niels Bohr contemplated the possibility that nuclear electrons not only did not behave as ordinary fermion but also interacted in a way that violated energy and momentum conservation, as was required to explain the continuous spectrum of $\beta$-decay. Far from considered crazy, Bohr's idea made it to the mainstream realm of textbooks: {\it ``This would mean that the idea of energy and its conservation fails in dealing with processes involving the emission or capture of nuclear electrons. This does not sound improbable if we remember all that has been said about peculiar properties of electrons in the nucleus.''}\cite{book_gamow}

\subsubsection{Pauli's Neutron Hypothesis}

Wolfgang Pauli came to the rescue in December, 1930, when he raised the hypothesis that there was a third constituent inside the nucleus. In his notorious letter addressed to the participants of a nuclear physics conference in T\"ubingen, Germany, included in summarized form below, Pauli postulated the existence of what he called a `neutron,' a fermion with no charge which interacted very weakly with matter and weighed less than 1\% of the proton mass. The presence of the neutron saved the spin-statistics theorem (e.g. $^{14}$N$=14p+7e^{-}+7$`$\nu$', a boson) and explained the apparent violation of energy conservation in $\beta$-decay. He assumed that $^A\hspace*{-0.8mm}Z\to ^A\hspace*{-1mm}(Z+1)+e^{-}+$`$\nu$' was the correct physical process and, since the final state contained three bodies, not only was the electron energy spectrum continuous, but the maximal electron energy was guaranteed to be always less than the parent--daughter mass-difference. It is, perhaps, also noteworthy to mention that Pauli did not publish his idea until several years later. He was worried about the fact that he could be postulating a new particle, whose existence would never be verified experimentally, in order to save a handful of theoretical principles\ldots

\framebox{
\begin{minipage}[t]{0.9\textwidth}
Dear Radioactive Ladies and Gentlemen, 

I have come upon a desperate way out regarding the wrong statistics of the $^{14}$N and $^6$Li nuclei, as well as the continuous $\beta$-spectrum, in order to save the ``alternation law'' statistics and the energy law. To wit, the possibility that there could exist in the nucleus electrically neutral particles, which I shall call ``neutrons,'' and satisfy the exclusion principle...
The mass of the neutrons should be of the same order of magnitude as the electron mass and in any case not larger than 0.01 times the proton mass. The continuous $\beta$-spectrum would then become understandable from the assumption that in $\beta$-decay a neutron is emitted along with the electron, in such a way that the sum of the energies of the neutron and the electron is constant\ldots

For the time being I dare not publish anything about this idea and address myself to you, dear radioactive ones, with the question how it would be with experimental proof of such a neutron, if it were to have the penetrating power equal to about ten times larger than a $\gamma$-ray.

I admit that my way out may not seem very probable {\it a priori} since one would probably have seen the neutrons a long time ago if they exist. But only the one who dares wins, and the seriousness of the situation concerning the continuous $\beta$-spectrum is illuminated by my honored predecessor, Mr Debye who recently said to me in Brussels: ``Oh, it is best not to think about this at all, as with new taxes." One must therefore discuss seriously every road to salvation. Thus, dear radioactive ones, examine and judge. Unfortunately, I cannot appear personally in T\"ubingen since a ball\ldots in Z\"urich\ldots makes my presence here indispensible\ldots . 

Your most humble servant, W.~Pauli
 
{\it Adapted summary of an English Translation to Pauli's letter dated December 4, 1930, from Ref.~\refcite{book}.}

\end{minipage}}

\subsubsection{The Neutron and the Neutrino, the Muon and the Pion}

A few of years later, Chadwick discovered a neutral nuclear constituent. By studying the properties of the neutral radiation $n$ emitted in the process $^9$Be$+\alpha\to^{12}$C$+n$, he found out that this so-called neutron was a deeply penetrating neutral particle slightly {\sl heavier} than the proton, quite distinct from $\gamma$-rays. This was not the neutron postulated by Pauli, but a different particle all together. Given the fact that Chadwick's neutron was much heavier than Pauli's one, Fermi renamed Pauli's neutron the `neutrino.'\footnote{In italian, the suffix `ino' is used to represent the diminutive --- `neutrino'~=~`small neutron.'}

More than baptizing the neutrino, Fermi wrote down, in 1934, a new quantum mechanical description of the weak interactions. He postulated that $\beta$-decay was mediated by the decay process $n\to pe^{-}\bar{\nu}_e$,\footnote{I will use modern notation for neutrinos and antineutrinos henceforth, in order to avoid more confusion.} which was described by the following four-fermion interaction:\footnote{Again, I use modern field theory notation to avoid confusion.}
\begin{equation}
\frac{G_F}{\sqrt{2}} \left(\bar{n}\Gamma_N p\right) \left(\bar{\nu}_e\Gamma_Le\right)+H.c.,
\label{fermi_theory}
\end{equation}
where $G_F$ is a dimensionful constant that characterizes the strength of the interaction (the Fermi constant), and $\Gamma_{N,L}$ are linear combinations of ``gamma matrices'' $1,\gamma_5,\gamma_{\mu},\gamma_{\mu}\gamma_5,\sigma_{\mu\nu}$, which were to be determined by more precise measurements of weak interaction processes. With this new understanding of the $\beta$-decay process, a much improved picture of the nucleus was built. Nuclei were built out of nucleons (neutrons and protons) --- $^AZ=Zp+(A-Z)n$, a concept which correctly explained the magnetic moment of the nuclei, and allowed one to determine, correctly, whether a given nucleus was a boson or a fermion (e.g. $^{14}$N$=7p+7n$, a boson).

Eq.~(\ref{fermi_theory}) not only provided a mathematical description of nuclear beta decay, but also allowed one to compute the cross-section for other related physical processes, including
\begin{equation}
\bar{\nu}_e + p \to e^+ + n, 
\label{nu_p_collision}
\end{equation}
which was crucial, many years later, to determine whether neutrinos could be experimentally observed. I will return to this issue shortly.

The importance of the Fermi theory and our understanding of the weak interactions increased tremendously in the years that followed. In 1936, a new elementary particle, the $\mu$ `meson' was discovered in cosmic ray experiments. This particle, now known as the muon (a lepton), was first confused with the pion, postulated by Hideki Yukawa to be the mediator of the interaction responsible for binding protons and neutrons inside the nucleus. This particular issue was resolved theoretically in 1947 by Marshak and Bethe with the ``two meson hypothesis.'' Their proposal was later confirmed to be correct when pions were first observed in cosmic ray experiments, also in 1947. As far as neutrinos are concerned, it is important to note that it was established that
\begin{eqnarray}
\pi^+\to \mu^+\nu_{\mu}, 
\label{pion_decay} \\
\mu^+\to e^+\nu_{e}\bar{\nu}_{\mu},
\label{muon_decay}
\end{eqnarray}
and that both the pion and muon decay rates were characterized by the same interaction strength which was found in nuclear decay processes, Eq.~(\ref{fermi_theory}). For example, the four-fermion operator 
\begin{equation}
\frac{G_F}{\sqrt{2}} \left(\bar{\nu}_{\mu}\Gamma \mu\right) \left(\bar{\nu}_e\Gamma e\right)+H.c.,
\label{muon_lagrangian}
\end{equation}
mediates Eq.~(\ref{muon_decay}). The fact that $G_F$ in Eq.~(\ref{fermi_theory}), is the same as $G_F$ in Eq.~(\ref{muon_lagrangian}) was the first indication that the weak interactions were a {\sl universal phenomenon}, much like electromagnetism, and not an exclusive property of nuclear systems.

A very long, eventful time passed between Fermi's formulation of the weak interactions and  the confirmation that Eq.~(\ref{fermi_theory}) was correct beyond any reasonable doubt plus the experimental  determination of the nature of the $\Gamma_N$'s. I refer readers to, e.g., Ref.~\refcite{book} and references therein for a detailed description. Of particular importance were precision studies of the energy spectrum of electrons emitted in muon decay (so-called Michel electrons) and searches for the helicity suppressed $\pi^+\to e^+\nu_e$ decay, finally observed for the first time in 1958. A couple of years before that, weak interactions caused a significant amount of commotion in the community with the hypothesis, by Lee and Yang, and subsequent experimental confirmation, first by Wu {\it et al}, and a little later by Garwin, Lederman, and Weinrich, and by Friedman and Telegdi, that parity was maximally violated in weak processes. The whole issue was finally settled in the same year, thanks to theoretical developments by Marshak and Sudarshan and Feyman and Gell-mann, which argued in favor of the now well known $V-A$ (or $\gamma_{\mu}-\gamma_{\mu}\gamma_5$) structure of the weak interactions.  

The purely left-handed structure of the leptonic charged current ($\bar{\nu}_e\gamma_{\mu}(1-\gamma_5)e$), combined with the fact that neutrino masses were known to be much smaller than the electron mass, allowed for a very compact description of the neutrino. A free, massive spin 1/2 particles is characterized, not surprisingly, by its mass $m$, its total spin $s$ the value of its spin projection in a specific direction, $s_z$. One particularly useful direction for measuring the particle's spin is its direction of motion $\hat{p}$, in which case $s_z$ is called the particle's handedness, or helicity. If $\vec{s}\cdot\hat{p}|m,s\rangle=+1/2|m,s\rangle$ ($-1/2|m,s\rangle$), the particle is said to be right-handed (left-handed). The helicity is not reference-frame independent in the case of massive particles. This is easy to see. If the particle is massive, one can always choose to change into a reference frame moving along its direction with a larger velocity. In this case, $\hat{p}\leftrightarrow-\hat{p}$, while $\vec{s}\leftrightarrow\vec{s}$, such that, say, a right-handed state is now viewed as a left-handed one. 

In the case of massless fermions, however, the helicity is a good quantum number, independent on the reference frame. The smallest number of degrees of freedom a massless fermion field can describe is, hence, two: a left-handed (or right-handed) fermion, plus its CPT-transform, a right-handed (or left-handed) antifermion.\footnote{This is to be contrasted to the case of charged, masssive fermions which require four degrees of freedom (e.g., the left-handed electron and its CPT-transform, the right-handed positron, plus its ``Lorentz transform'' the right-handed electron and CPT-transform, the left-handed positron). See Ref.~\refcite{willen} for a more detailed discussion of this point.} Because the weak-interactions are purely left-handed, this has proven, so far, to be enough to describe the neutrinos. 

One of the most dramatic manifestations of this phenomenon is the fact that muons emitted in pion decay at rest are virtually 100\% polarized. This is a consequence of the fact that the weak interactions are purely left-handed, and that the neutrino mass is, at most, tiny:
\begin{eqnarray}
&\pi^+\to\mu^+_L\nu_{\mu,L}~~~~\rm ALWAYS; \\
&\downarrow(P) \nonumber \\
&\pi^+\to\mu^+_R\nu_{\mu,R}~~~~\rm NEVER; \\
&\downarrow(C) \nonumber \\
&\pi^-\to\mu^-_R\bar{\nu}_{\mu,R}~~~~~\rm ALWAYS; \\
&\downarrow(P) \nonumber \\
&\pi^-\to\mu^-_L\bar{\nu}_{\mu,L}~~~~~\rm NEVER. 
 \end{eqnarray}
In summary, in the absence of neutrino masses and new interactions beyond $SU(2)_L$, right-handed neutrinos and left-handed antineutrinos can never be produced or detected. This is equivalent to saying they don't exist.

Before proceeding, I'll mention that the neutrino helicity was first determined in 1958 by Goldhaber {\it et al.} in a very elegant experiment. They produced neutrinos via $e^-+^{152}$Eu$\to\nu_e+^{152}$Sm$^*(J=1)$. The excited state of samarium decayed very quickly by emitting a  $\gamma$-ray, $^{152}$Sm$^*(J=1)\to^{152}$Sm$(J=0)+\gamma$. Conservation of energy and momentum forces the photon and the neutrino to be back-to-back, while conservation of angular momentum correlates the neutrino helicity to the photon polarization. Goldhaber {\it et al.} established that the photons produced in these processes were, within errors, 100\% polarized, and that the neutrinos were purely left-handed.

\subsubsection{Direct Detection of (Anti)Neutrinos}

While the neutrino was postulated to exist in 1930 and widely accepted to be a real particle shortly thereafter, it was not until the 1950's that the necessary means for detecting neutrinos became available. As mentioned earlier, Fermi theory allowed one to compute the neutrino--nucleon cross section, and hence estimate the neutrino flux and detector size required to statistically observe neutrino-induced events. 

In the early 1950's ``Project Poltergeist,'' headed by Frederick Reynes and Clyde Cowan, got started. Its mission was to detect neutrinos. It took advantage of the technological developments which took place in the 1940's in order to develop the fission bomb. Curiously enough, one of the first proposals consisted of measuring the (very high) instantaneous flux of antineutrinos emitted {\sl during} an atomic bomb explosion (remember that nuclear tests in the american desert were taking place at the time!). One of the down sides of this particular setup included the fact that the experiment needed to be located rather close to the site of the explosion --- not the most stable of places to house a detector!    

This rather original idea --- never implemented --- was substituted with the idea to measure the antineutrino flux produced by nuclear power reactors. Nuclear reactors were also a ``bi-product'' of the bomb, and were used to enrich heavy isotope samples. While the instantaneous antineutrino flux from nuclear reactors was much less than the one generated during a nuclear explosion, the integrated flux could be much larger --- and the experimental conditions were, of course, much more sensible. 

In order to successfully detect neutrino-induced interactions, it was crucial to separate these from background events due to cosmic rays, radioactivity in the detector and surrounding material, etc. This was accomplished successfully in the following way. Antineutrinos interact with protons as depicted in Eq.~(\ref{nu_p_collision}). The daughter positron quickly annihilates with a near-by electron
\begin{equation}
e^+e^-\to\gamma\gamma,
\end{equation}
and the photon energy is detected in some scintillating environment (furthermore, the total deposited energy is related to the positron energy, which is related to the incoming antineutrino energy). While the positron is being annihilated, the recoil neutron random walks inside the detector and is absorbed, after a well defined characteristic time, emitting  $\gamma$-rays with well-defined energy. The coincidence between the first signal (due to the elelctron-positron annihilation) and second one, a well-defined amount of time afterwards, was crucial to the success of Project Poltergeist, and is used to this very day to study antineutrinos produced in nuclear reactors. 
 
 The first hint of $\bar{\nu}_e+p$ scattering events was obtained in 1953, in a one cubic meter scintillator tank near the Hanford reactor site. This setup was swamped with cosmic ray-induced background events, and obtained a two-sigma evidence for inverse beta-decay ($\bar{\nu}_e+p\to e^++n$). Definitive evidence was only obtain in 1956 (final results presented in 1960) with a larger detector next to the Savannah River reactor site. The key changes were related to improved neutron detection techniques and much better cosmic ray veto. Reynes was awarded a Physics Nobel Prize for ``pioneering experimental contributions to lepton physics'' in 1995.
 
 It is relevant to note that a different technique for observing antineutrinos was tried, without success, in ``parallel'' with Project Poltergeist. In 1955, a radio-chemical experiment, led by Ray Davis, located next to a nuclear reactor site failed to observe ``inverse chlorine decay,'' {\it i.e.},
 \begin{equation}
 \bar{\nu}_e+^{37}\hspace*{-1mm}{\rm Cl}\to e^-+^{37}\hspace*{-1mm}{\rm Ar}
 \label{anti_davis}
 \end{equation}
 does not happen with a measurable rate. This null result can be interpreted as evidence that neutrinos and antineutrinos are distinct particles.\footnote{Or that they have a very tiny (or no) mass.} We currently interpret the apparent impossibility of Eq.~(\ref{anti_davis}) as a consequence of a conservation law --- lepton number conservation. Neutrinos and negatively charged leptons are assigned lepton number +1, while antineutrinos and positively charged leptons are assigned lepton number $-1$. Eq.~(\ref{anti_davis}) violates lepton number by two units. A similar setup was eventually used, very successfully, to study an intense (natural!) source of {\sl neutrinos} --- the Sun.  
 
\subsubsection{And Then There Were Two (Then Three)}

As alluded to earlier, the left-handed nature of the weak interactions plus CPT-invariance only require the existence of one left-handed neutrino, and one right-handed antineutrino. However, soon after the discovery of the muon, it was hypothesized that there were two distinct neutrino `flavors.' In modern language, the question which was to be addressed experimentally in the early 1960's was: are muon-type neutrinos $\nu_{\mu}$ different particles from electron-type neutrinos $\nu_e$? 

\begin{figure}[ht]
\centerline{\epsfig{width=0.8\textwidth, file=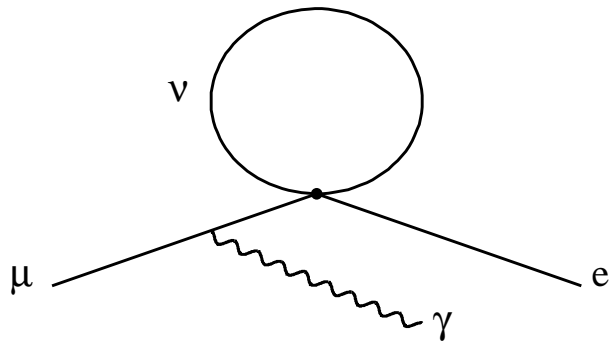}}
\caption{One of the Feynman diagrams contributing to $\mu\to e\gamma$ at the one-loop level in the Fermi theory, assuming that $\nu_{\mu}$ and $\nu_{e}$ are the same particle. \label{meg_effective}}
\end{figure}

The reason to suspect that this is the case is the fact that, while the muon decays into an electron plus two neutrinos (according to the above mentioned conservation of lepton number, the muon decays into an electron plus a neutrino plus an antineutrino, as depicted in Eq.~(\ref{muon_decay})), it did not decay into an electron plus a photon, {\it i.e.}, $\mu^{\pm}\to e^{\pm}\gamma$ has never been observed. Currently, we have bound $Br(\mu^{+}\to e^{+}\gamma)<1.2\times 10^{-11}$ at the 90\% confidence level.\cite{PDG} If, however, $\nu_{\mu}$ and $\nu_e$ were the same particle, $\mu^{\pm}\to e^{\pm}\gamma$ would happen at one-loop order, as depicted in Fig.~\ref{meg_effective}.

The absence of $\mu^{\pm}\to e^{\pm}\gamma$ can also be reinterpreted in terms of a new conservation law --- conservation of individual lepton number. If one postulates that the electron has electron-number +1 while the muon has muon-number +1, electron-number and muon-number conservation forbid $\mu^{\pm}\to e^{\pm}\gamma$. Furthermore, there must be at least two distinct neutrino ``flavors:'' the $\nu_e$ with electron-number +1, produced together with electrons in, say, nuclear $\beta$-decay, and the $\nu_{\mu}$ with muon-number +1, produced together with muons in, say, pion decay (Eq.~(\ref{pion_decay})).

That muon-type and electron-type neutrinos are indeed different particles was experimentally established in 1962, in an effort led by Leon Lederman, Jack Steinberger, and Melvin Schwarts at the Brookhaven National Laboratory. They were awarded the Physics Nobel Prize in 1988 for their discovery (``for the neutrino beam method and the demonstration of the doublet structure of the leptons through the discovery of the muon neutrino''). The key to the experiment was the development of the first neutrino beam. Muon neutrino beams are relatively straightforward to produce. By colliding protons on a target at large enough energies, one produces a large flux of pions. The charged pions eventually decay, more than 99\% of time, into muons and neutrinos. One can aim the pions (and hence the muons and the neutrinos) into a beam-dump (a large amount of material which will absorb all charged leptons and hadrons in the beam) and place a detector on the other end. The detector will be traversed by a flux of only neutrinos, given that all other particles in the original beam will have been absorbed by the beam dump. 

Because the neutrinos are mostly produced by charged pion decay, their vast majority will be of the muon-type.\footnote{There other components to the neutrino beam. Kaons, also copiously produced by proton--target interactions, decay into both electron-type and muon-type neutrinos, while some of the muons produced by pion decay also decay in flight, emitting both electron-type and muon-type neutrinos. The $\nu_e$ contamination level depends on the energy of the incident proton beam, the ``pion-focusing'' mechanism, and the length of the decay tunnel. While not too relavant for the Lederman-Steiberger-Schwartz experiment, these are all crucial issues for modern neutrino beam experiments.} The question one wishes to address is: do these neutrinos, when interacting inside the detector, produce muons or electrons?, {\it i.e.}, is 
\begin{eqnarray}
&\nu_{\mu}+X\to \mu+Y & {\rm and/or} \label{numu_to_mu} 
\\
&\nu_{\mu}+X\to e+Y &
\label{numu_to_e}
\end{eqnarray}
allowed ($X$ and $Y$ are irrelevant initial and final states)? Eq.~(\ref{numu_to_e}), while enjoying a larger available phase space (the electron is 200 times lighter than the muon), violates muon-number and electron-number conservation. It was not observed in 1962, while Eq.~(\ref{numu_to_mu}) was observed, at a rate consistent with expectations from weak-interactions.
 
From 1974 to 1977, experiments led by Martin Perl (Physics Nobel Prize 1995), revealed the existence of a third lepton, the tau ($\tau$). It was immediately recognized that, along with the $\tau$, there should be a third neutrino flavor $\nu_{\tau}$. Hard evidence for the existence of $\nu_{\tau}$ was only obtained much later. First indirectly, when precision measurements of the line-shape of the $Z^0$-boson, performed at LEP, revealed the existence of an invisible $Z^0$-boson width consistent with the standard model prediction as long as there were three neutrino species.\cite{LEP_three} More indirect evidence can be obtained from detailed calculations of the relic abundance of (mostly) $^4$He, which depends on the expansion rate of the Universe around the time of Big-Bang nucleosynthesis ($T\sim1$~MeV).\cite{BBN} These computations reveal, with large error bars, that the number of relativistic degrees of freedom present at the time of  Big-Bang nucleosynthesis is consistent with the existence of three neutrinos.
Evidence for tau-type neutrinos similar to the one obtained by the Lederman-Steiberger-Schwartz experiment was only obtained in 2001, when the DONUT (``Direct Observation of NU Tau'') experiment at Fermilab managed to record a handful of $\nu_{\tau}+X\to\tau+Y$ events.\cite{DONUT}
 
To conclude this subsection, it is interesting to appreciate that most of the conclusions obtained in the paragraphs above are known today to be, at a more fundamental level, incorrect. We have learned from neutrino experiments that the conservation of individual lepton number is strongly violated. However, individual lepton number violating effects can only be observed under rather special circumstances, which will be discussed in the upcoming sections, and the rate for $\mu\to e\gamma$ is not zero but severely suppressed if only mediated by electroweak interactions. The reason for this is simple.  $\mu\to e\gamma$ is a leptonic example of flavor changing neutral current processes, which are known to be GIM suppressed. In a nutshell, such processes are not mediated at the tree level by $Z^0$-boson exchange, while higher order effects are suppressed by the unitarity of the fermion mixing matrix, in such a way that $A(\mu\to e\gamma)\propto \Delta m^2$, where $\Delta m^2$ are neutrino mass-squared differences. Given what we have learned about leptonic mixing and neutrino mass-squared difference, the rate for $\mu\to e\gamma$ is absurdly small:\cite{mutoegamma}
\begin{equation}
Br(\mu \to e \gamma )=\frac{3\alpha}{32 \pi}\left|\sum_i 
U_{\mu i}^*U_{ei} \frac{\Delta m_{1i}^2}{M_W^2}\right|^2 \lesssim 10^{-56},
\end{equation}
where $U_{\alpha i}$ are the elements of the leptonic mixing matrix, while $\Delta m^2_{1i}$, $i=2,3$ are the neutrino mass-squared differences. For this reason, searches for rare muon processes, including $\mu\to e\gamma$, $\mu\to e^+e^-e$ and $\mu+^AZ\to e+ ^AZ$ ($\mu$-$e$--conversion in nuclei) are considered ideal laboratories to probe effects of new physics at or slightly above the electroweak scale.\cite{Kuno-Okada}
 
\subsection{Neutrino Puzzles}

As the neutrinos and their properties were being uncovered and explored, a series of anomalies popped-up. These started as curious experimental results which disagreed with theoretical expectations and were quickly dismissed by most of the community for various reasons. In time, a handful of these evolved into well-respected puzzles (or problems), and all but one (which is discussed in Sec.~\ref{sec_lsnd}) have been resolved in a most surprising way. Here, I'll describe the puzzles that lead to the discovery of neutrino flavor change and, ultimately, to the current understanding that neutrinos have mass and that leptons mix strongly. I'll deviate significantly from any historical timeline, and will steer away from discussions about the history of neutrino oscillations.\cite{review_history,book}

\subsubsection{Solar Neutrinos}

Soon after it was understood that the Sun burns via nuclear fusion, it was also appreciated that the Sun was a ``$\nu_e$ factory,'' {\it i.e.}, the thermonuclear reactions that take place inside the Sun's core produce both photons and neutrinos. We now have enough evidence to believe that most of the Sun's energy is produced by proton--proton fusion, a process through which, roughly speaking $p+p+p+p\to^4$He$+e^++e^++\nu_e+\nu_e+\gamma$'s. The $pp$-chain is depicted in Table~\ref{pp-table}.

\begin{table}	
\tbl{Nuclear reactions responsible for producing almost all of the Sun's energy and the different ``types'' of solar neutrinos (nomenclature): $pp$-neutrinos, $pep$-neutrinos, $hep$-neutrinos, $^7$Be-neutrinos, and $^8$B-neutrinos. `Termination' refers to the fraction of interacting protons that participate in the process. \label{pp-table}}
{\begin{tabular}{cccc}
\hline \\
Reaction & Termination  & Neutrino Energy & Nomenclature \\ 
 & (\%) & (MeV) & \\
\hline \\
$p+p\to ^2$H$+e^++\nu_e$ & 99.96 & $<0.423$ & $pp$-neutrinos \\ \\
$p+e^-+p\to ^2$H$+\nu_e$ & 0.044 & 1.445 & $pep$-neutrinos \\ \\
$^2$H$+p\to^3$He$+\gamma$ & 100 & -- & -- \\ \\
$^3$He$+^3$He$\to^4$He$+p+p$ & 85 & -- & -- \\ \\
$^3$He$+^4$He$\to^7$Be$+\gamma$ & 15 & -- & -- \\ \\
$^7$Be$+e^-\to^7$Li$+\nu_e$ & 15 & $\begin{array}{c} 0.863 (90\%)\\ 0.386 (10\%) \end{array}$ & $^7$Be-neutrinos \\ \\
$^7$Li$+p\to^4$He$+^4$He &  & -- & -- \\ \\
$^7$Be$+p\to^8$B$+\gamma$ & 0.02 & -- & -- \\ \\
$^8$B$\to^8$Be$^*+e^++\nu_e$ & & $<15$ & $^8$B-neutrinos \\ \\
$^8$Be$\to^4$He$+^4$He & &  --  & -- \\ \\
$^3$He$+p\to^4$He$+e^++\nu_e$ & 0.00003 & $<18.8$ & $hep$-neutrinos \\ 
\hline
\end{tabular}}
\begin{tabnote}
Adapted from Ref.~\refcite{bahcall_book}. Please refer to Ref.~\refcite{bahcall_book} for a more detailed explanation.
\end{tabnote}
\end{table}

\begin{figure}[ht]
\centerline{\epsfig{width=0.8\textwidth, angle=-90, file=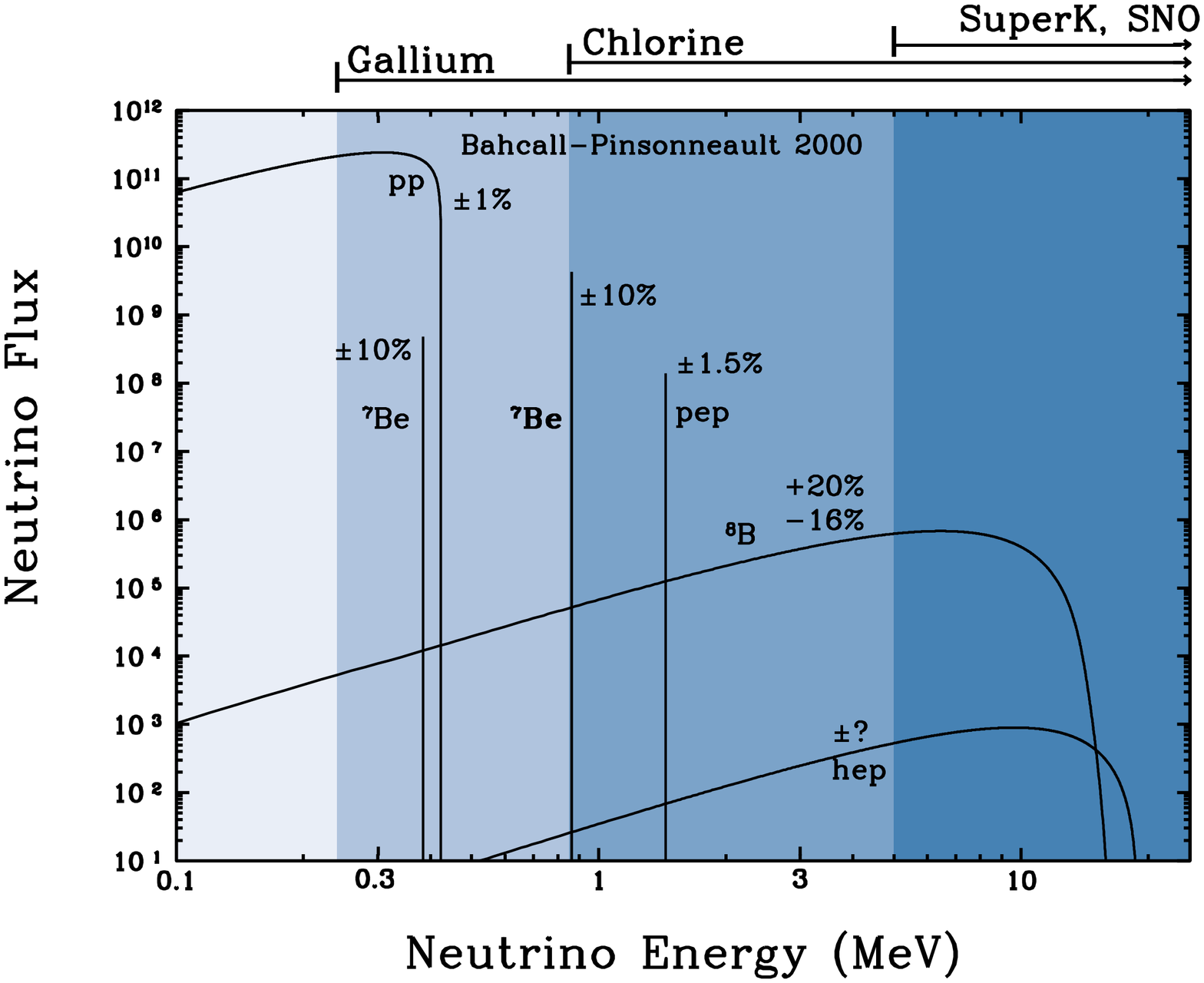}}
\caption{The predicted solar neutrino energy spectrum.  The figure shows the energy spectrum of solar neutrinos predicted by the most recent version of the standard solar model. For continuum sources, the neutrino fluxes are given in number of neutrinos cm$^{-2}$s$^{-1}$ MeV$^{-1}$ at the Earth's surface.  For line sources, the units are number of neutrinos cm$^{-2}$s$^{-1}$. Total theoretical uncertainties are shown for each source. From http://www.sns.ias.edu/$\sim$jnb/. 
\label{solar_fluxes}}
\end{figure}

In the 1940's and 1950's, the crucial question was whether these solar neutrinos could be detected here on Earth. The detection of solar neutrinos would serve, for example, as evidence that the Sun indeed obtained its energy from nuclear fussion processes. The challenge was two-fold. First of all, one needed to compute accurately enough the ``standard solar model'' expectations for the solar neutrino fluxes. Current standard solar model\cite{BP04} results for the solar neutrino fluxes (separated per ``type,'' see Table~\ref{pp-table}) are depicted in Fig.~\ref{solar_fluxes}. Second of all, given such a prediction, it was necessary to determine whether one could conceive of an experiment sensitive to the expected fluxes. 
 
To make a long story short, one of the most impressive achievements of nuclear experimental physics was obtained in 1964, when an experiment locate at the Homestake mine in South Dakota, led by Ray Davis, obtained evidence for a solar electron-type neutrino flux, roughly consistent with solar model expectations. The experiment detected electron-type neutrinos via inverse nuclear $\beta$-decay
\begin{equation}
\nu_e+^{37}\hspace*{-1mm}{\rm Cl}\to e^-+^{37}\hspace*{-1mm}{\rm Ar},
\end{equation}
using a technique not dissimilar to the one which failed to see electron-type {\sl anti}neutrinos from reactors. The experiment consisted of a very large tank containing a chlorine compound. The tank was ``searched'' periodically for argon atoms, which were then detected by the decay of radioactive $^{37}$Ar. In order to appreciate how challenging the experiment was, in several cubic meters of the chlorine compound, several argon atoms were detected --- every {\sl month}! Davis was awarded the 2002 Physics Nobel Prize ``for pioneering contributions to astrophysics, in particular for the detection of cosmic neutrinos.''

The Homestake experiment continued to measure the solar neutrino flux for over thirty years. It was followed by two different types of experiments. The Kamiokande experiment (start date 1985) was a very large water Cherenkov experiment, designed to look for proton decay. It also managed to detect neutrinos from the Sun, via elastic neutrino--electron scattering: $\nu_e+e^-\to\nu_e+e^-$. The Cherenkov light emitted by the recoil electron is observed and used to reconstruct the electron energy and direction, which is correlated to the incoming neutrino energy and direction. The GALLEX (Italy) and SAGE (USSR/Russia) experiments (start date 1991/1990) were, similar to Homesake, also radiochemical experiments, and detected neutrinos via inverse nuclear $\beta$ decay of gallium: $\nu_e+^{71}$Ga$\to e^-+^{71}$Ge. As in the Homestake experiment, chemical techniques were used to isolate and count the number of radioactive $^{71}$Ge atoms produced by the neutrino reaction.

The three different types of experiments provided complementary information. The water Cherenkov experiment was capable of detecting neutrinos in real time, and determine their energy. They were also the first to correlate the incoming neutrino direction with the position of the Sun in the sky. On the other hand, the water Cherenkov technique is only sensitive to very ``high energy'' solar neutrinos, and can only see the $^8$B neutrinos. The radiochemical experiments do not have any capability to distinguish the energy of the incoming neutrinos, but have lower energy detection thresholds. The chlorine experiment was sensitive to neutrino energies higher than $\sim1$~MeV, while the gallium experiments were also sensitive to the $pp$-neutrinos. The different energy thresholds for the different detection techniques are depicted in Fig.~\ref{solar_fluxes}.

\begin{figure}[ht]
\centerline{\epsfig{width=0.8\textwidth, angle=-90, file=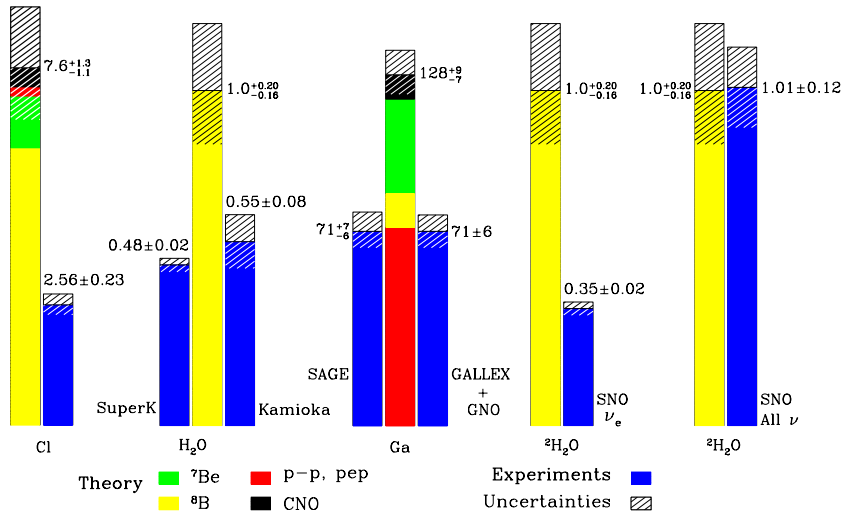}}
\caption{Predictions of the standard solar model and the total observed rates in the six solar neutrino experiments: chlorine, SuperKamiokande,  Kamiokande, GALLEX, SAGE, and SNO. The  model predictions are color coded with different colors for the different predicted neutrino components. For both the experimental values and the predictions, the  1 sigma uncertainties are indicated by cross hatching. From http://www.sns.ias.edu/$\sim$jnb/. 
\label{solar_results}}
\end{figure}

After the first Homestake results were published in the mid 1960's, a salient feature of the solar neutrino data became apparent: the measured neutrino flux was statistically smaller than theoretical expectations. These original results were quickly dismissed due to problems with the experiment and/or with the theoretical computations of the solar neutrino flux.  The ``solar neutrino anomaly'' did not go away with the advent of the Kamiokande, GALLEX, and SAGE data --- all three experiments confirmed the solar neutrino deficit observed by Davis's experiment. Fig.~\ref{solar_results} depicts the solar neutrino flux measured by the different experiments, along with the expectations of the standard solar model.

A combined analysis of Homestake, SAGE, GALLEX, and Kamiokande data revealed that plausible astrophysical solutions to the solar neutrino problem were safely ruled out. Several logical explanations for the deficit remained, and more experimental information, which was provided first by the Super-Kamiokande experiment, later by the SNO experiment, and finally by the KamLAND experiment, was required to sort things out. These will be presented later.

\subsubsection{Atmospheric Neutrinos}

In the 1980's, experiments started measuring  the flux of atmospheric neutrinos more precisely. These are neutrinos produced by the interactions of cosmic rays and the atmosphere. In more detail, cosmic rays (mostly protons) hit the atmosphere and produce a shower of mostly pions. The pions decay in flight into muons and muon-type neutrinos. The muons later decay into electrons, electron-type neutrinos and muon-type antneutrinos. In summary
\begin{eqnarray}
\pi^{\pm}\to\mu^{\pm}&+&\nu_{\mu}~({\rm or}~\bar{\nu}_{\mu}), \\
 &\searrow &  \nonumber \\
 & & e^{\pm} \nu_e~({\rm or}~\bar{\nu}_e)~\bar{\nu}_{\mu}~({\rm or}~\nu_{\mu}).
\end{eqnarray}
For low enough neutrino energies, the muon-flavor to electron-flavor ratio is expected to be $2:1$. This ration is expected to increase as the neutrino energy increases due to the fact that the fraction of muons that decays in flight decreases as the muon energy increases. 

Several different experiments were built in order to study atmospheric neutrinos. Among them, NUSSEX, Frejus, Soudan, and MACRO were all calorimeter-like detectors, while Kamiokande and IMB were water Cherenkov detectors, built originally to look for proton decay.\footnote{Kamiokande stands for `Kamioka Nucleon Decay Experiment.' Kamioka is the name of a mine in Japan.} The later two experiments studied the atmospheric neutrino flux because these were expected to provide one of the dominant sources of background for nucleon decay searches.

Unlike solar neutrinos, the flux of atmospheric neutrinos cannot be computed very accurately. Experiments measure theoretically robust quantities in order to study these neutrinos. One of these is the muon-type to electron-type flux ratio or the $R$-ratio, defined to be the  measured muon-type to electron-type flux ratio divided by its theoretical prediction. Results for $R$ are listed in Table~\ref{r-summary}, courtesy of Ref.~\refcite{sum_atm}.
\begin{table}
\tbl{Summary of $R$ measurements.\label{r-summary}}
{\begin{tabular}{lrrlr}
\hline
\multicolumn{1}{c}{Experiment} & 
\multicolumn{1}{c}{kt-yr} & 
\multicolumn{1}{c}{events} & 
\multicolumn{1}{c}{$R$ (data/MC)} &
\multicolumn{1}{c}{``material''} \\ 
\hline
IMB
& 7.7 & 610 & $0.54\pm.05\pm.11$ & water \\
Kamiokande
& 7.7 & 482 & $0.60^{+.06}_{-.05}\pm.05$ & water\\
Soudan-2
& 3.2 & $\sim$200 & $0.61\pm.15\pm^.05$ & iron\\
Fr\'{e}jus
& 2.0 & 200 & $1.00\pm.15\pm.08$ & iron \\
NUSEX
& 0.7 & 50 & $0.96^{+.32}_{-.28}$ & iron \\
\hline
\end{tabular}}
\end{table}

There are few characteristics of the results in  Table~\ref{r-summary} worth pointing out. First, three out of five results point to a value of $R$ statistically smaller than 1, which indicates that either the $\nu_{\mu}$ flux is smaller than expected, or that the $\nu_e$ flux is larger than expected. Second, the Soudan-2 result is more recent than the other four. In its absence, there is a clear correlation: calorimeter-like experiments obtained $R$ values consistent with one, while water Cherenkov detectors observed statistically smaller values.

Another part of this ``atmospheric neutrino anomay'' was the fact that the Kamiokande experiment provided more information than just $R$. It also measured the muon-type and electron-type neutrino fluxes as a function of the neutrino direction, and observed that, while the electron-type neutrino flux was roughly independent of whether the neutrinos were coming from above or for below, the muon-type neutrino flux was larger from above than from below.\cite{kamiokande_atm}
  
Any doubt about the reality of the atmospheric neutrino anomaly was erased after the Super-Kamiokande experiment first measured the atmospheric neutrino flux, as will be discussed later. Even before that, however, several physics explanations for the disappearance of muon-type neutrinos that traverse the entire Earth before being detected were explore. I summarize some of these below.

\subsubsection{Many Solutions, Super-Kamiokande, and SNO}

Both the solar and the atmospheric neutrino anomalies could be explained by postulating that the standard model description of neutrino production, propagation, and/or detection was incorrect. There were several logical explanations for the general features observed, namely the fact that only less than half the expected solar electron-type neutrinos produced in the Sun were actually detected, while a similar fraction of muon-type atmospheric neutrinos also disappeared.  Some of the logical possibilities include
\begin{enumerate}
\item Neutrinos have a finite lifetime, and decay into either other standard model particles or into new very light degrees of freedom (solar/atmospheric);
\item Neutrinos traversing large quantities of matter are absorbed (much, much) more efficiently than predicted by weak-interactions (solar/atmospheric);
\item Neutrinos have a small magnetic moment, and can be converted into either anti-neutrinos or ``right-handed neutrinos'' in the presence of intense magnetic fields. Electron antineutrinos are invisible to Homestake and the gallium experiments, and have a smaller cross-section for elastic scattering on electrons [Kamiokande], while ``right-handed'' neutrinos are virtually invisible (solar);
\item all kinds of exotica, including violations of the equivalent principle, violations of the unitary evolution of quantum mechanical states, violations of Lorentz invariance, etc (solar/atmospheric);
\item Neutrinos change flavor while they propagate, {\it i.e.}, a neutrino produced in a well-defined flavor $\nu_{\alpha}$ has a nonzero probability $P_{\alpha\beta}$ of being detected as a different flavor $\nu_{\beta}$, $\alpha,\beta=e,\mu,\tau$. If this is the case, the solar anomaly required $P_{ee}<1$, while the atmospheric anomaly required $P_{\mu\mu}<1$. Detailed analyses of the neutrino data require $P_{\alpha\beta}$ to depend on the distance travelled by the neutrino, and the neutrino energy (solar and atmospheric).
\end{enumerate}
A few of the naive solutions stated above were probably inconsistent with other data (like (2)), and failed to explain the neutrino data in detail ((3), for example, does not address the atmospheric anomaly). It is curious to note that several of them require that neutrinos have nonzero masses ((1) is the most obvious one).

In order to address which solution, if any, was correct, new experiments were required, and several were on the way. One was the Super-Kamiokande experiment, an improved and much larger version of the original Kamiokande experiment. Super-Kamiokande was designed not only to improve the experimental sensitivity to proton decay but also the determine whether the atmospheric neutrino anomaly was real (and to study it in more detail) and to provide a precise measurement of the $^8$B solar neutrino spectrum.

\begin{figure}[ht]
\centerline{\epsfig{width=0.5\textwidth, file=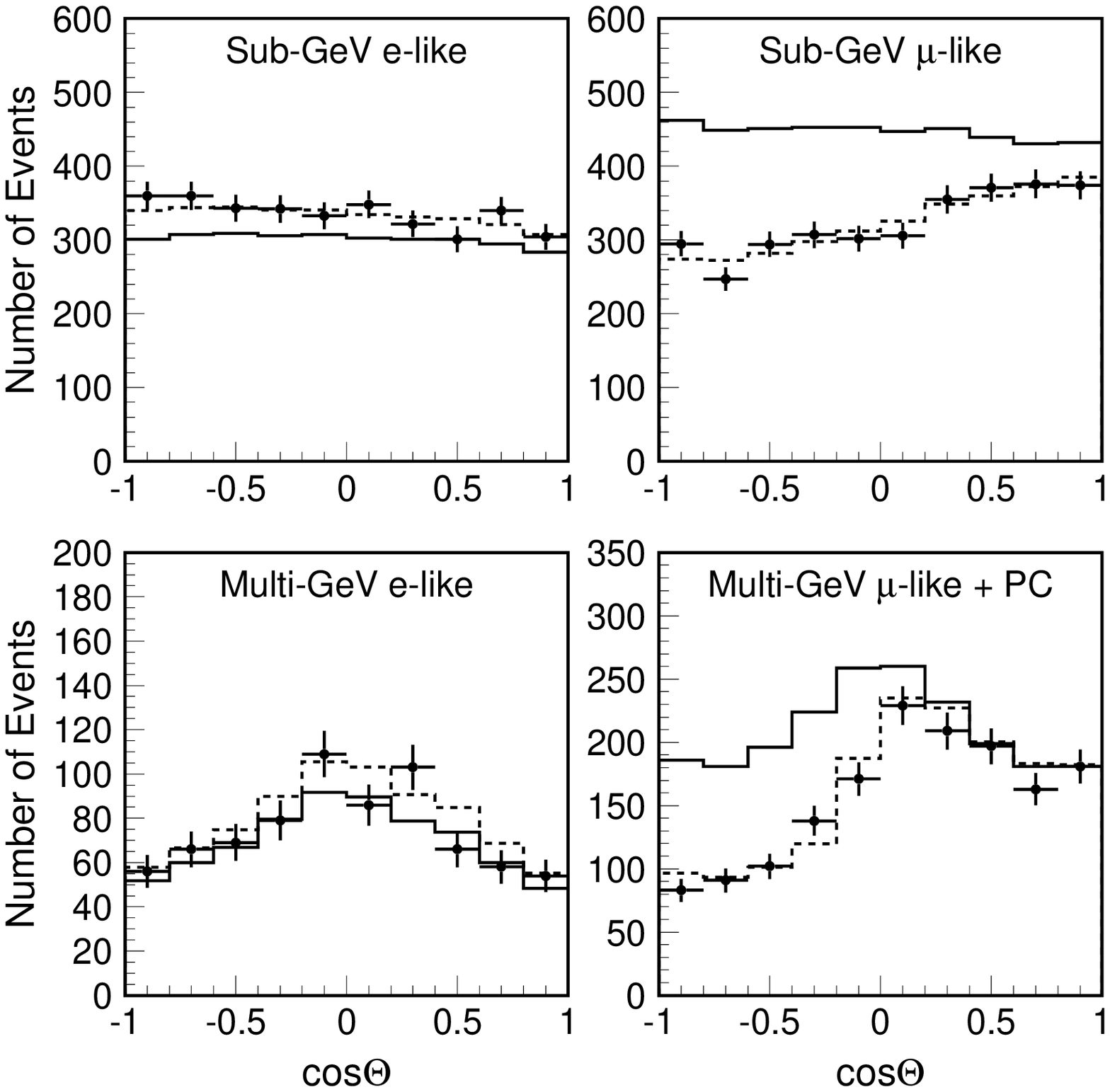}
\epsfig{width=0.5\textwidth, file=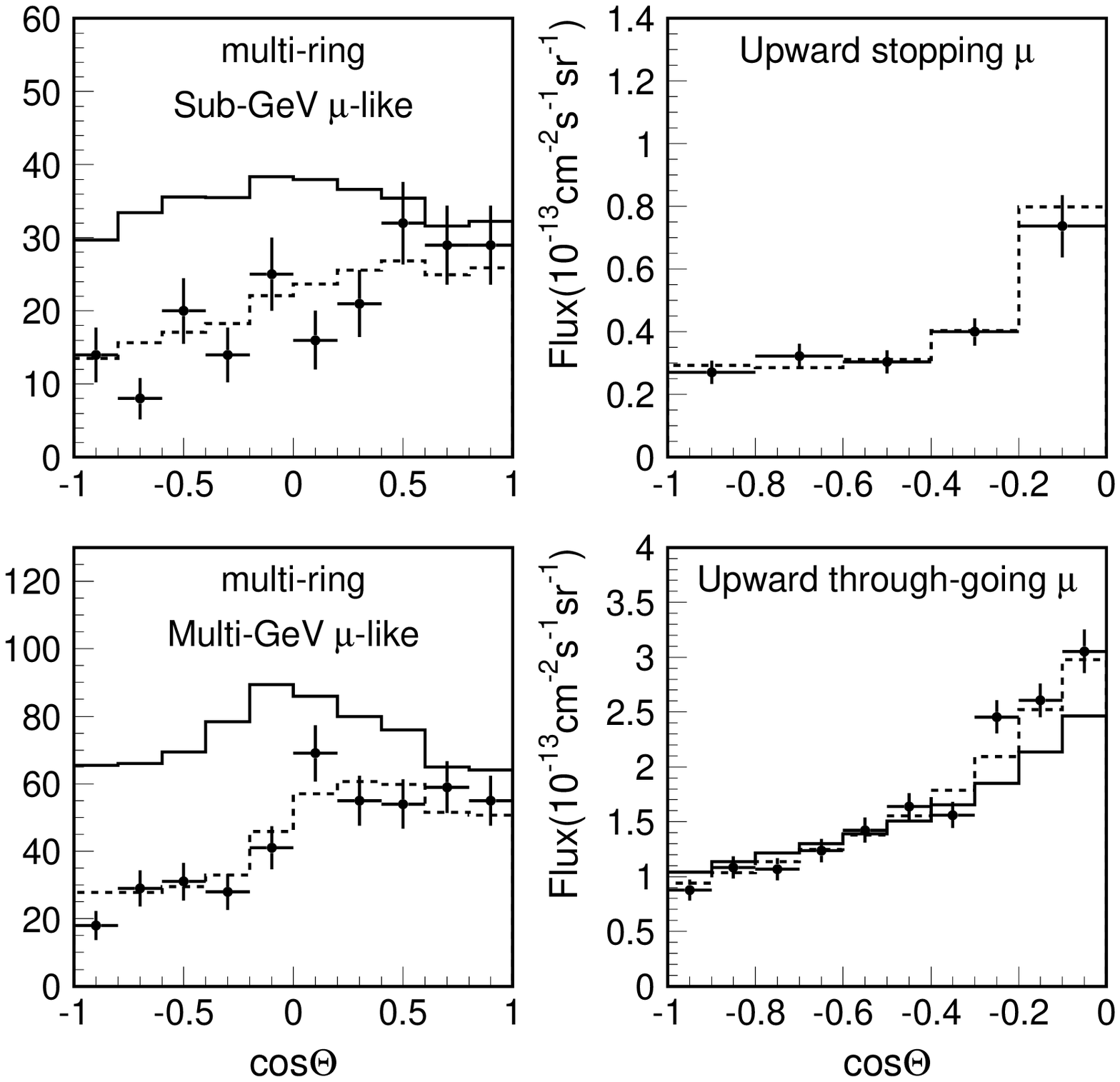}
}
\caption{Zenith angle distribution for fully-contained single-ring $e$-like and $\mu$-like events, multi-ring 
$\mu$-like events, partially contained events and upward-going muons. The points show the data and the solid
lines show the Monte Carlo events without neutrino oscillation.
 The dashed lines show the best-fit expectations for $\nu_\mu \leftrightarrow \nu_\tau$
 oscillations. From M.~Ishitsuka  [Super-Kamiokande Collaboration], hep-ex/0406076. \label{fig:superK_atm}}
\end{figure}

Arguably, no recent experiment has shone more light into our understanding of neutrinos than Super-Kamiokande. In the ``atmospheric sector,'' Super-Kamiokande provided evidence beyond any reasonable doubt that muon-type atmospheric neutrinos were indeed disappearing.\cite{SK_atm} It also established that the ``disappearance rate'' depends on the neutrino energy and baseline. The up-down  ratio of muon-type neutrinos at Super-Kamiokande is currently over 10 sigma away from one! The angular and energy dependency of the atmospheric neutrino flux is among the most remarkable particle physics results obtained in the past six years, and is depicted in Fig.~\ref{fig:superK_atm}.  

In the solar sector, Super-Kamiokande data confirmed the solar neutrino deficit (Fig.~\ref{solar_results}), but also provided evidence that the deficit was energy independent in the interval $\sim 5-10$~MeV. Super-Kamiokande also established that the solar neutrino flux did not depend anomalously on time (yearly, monthly, or daily).\cite{SK_solar}

The Sudbury Neutrino Observatory (SNO) was designed to perform independent measurements of the $^8$B solar neutrino flux.\cite{SNO} More specifically, SNO is a heavy water detector, and measures solar neutrino via three distinct processes:
\begin{eqnarray}
&\nu+^2\hspace*{-1mm}{\rm H} \to  p+p+\underline{e}^- & ~~~~~\nu_e~{\rm only}, \label{SNO_CC} \\
&\nu + e^- \to \nu+\underline{e}^- &  ~~~~~\nu_e + 0.15 ~\nu_{\mu,\tau}, \label{SNO_ES}\\
&\nu+^2\hspace*{-1mm}{\rm H}  \to \nu+p+\underline{n} & ~~~~~\nu_e+ \nu_{\mu,\tau}, \label{SNO_NC}
\end{eqnarray} 
where the underlining indicates the particle observed by SNO. Electrons produced by neutrino--electron scattering are separated from those produced by neutrino--deuteron scattering statistically, given the different kinematics of the different reactions. Neutron detection is done via different techniques, including detecting the photons emitted by neutron capture in deuteron and chlorine (these are also separated from the electron signals on a statistical basis), and placing $^3$He-filled neutron detectors inside the heavy water vessel. 

\begin{figure}[ht]
\centerline{\epsfig{width=0.8\textwidth, file=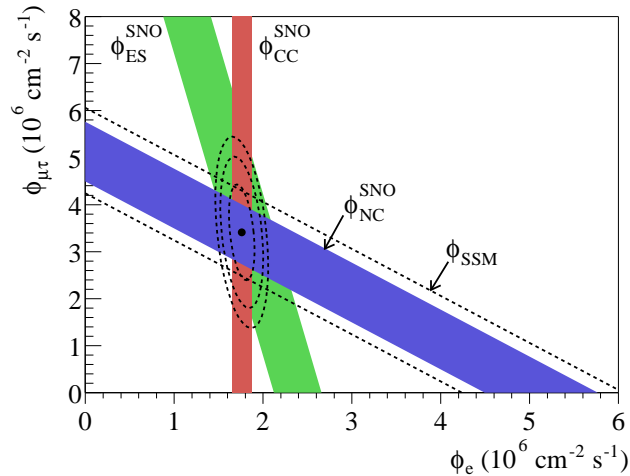}}
\caption{Flux of $^8$B solar neutrinos which are of the muon or tau-type, $\phi_{\mu\tau}$, versus
flux of electron neutrinos, $\phi_e$, deduced from the three neutrino reactions in SNO. The diagonal bands show the total 8B flux as predicted by the standard solar model (dashed lines) and that measured
with the NC reaction in SNO (solid band). The intercepts
of these bands with the axes represent the $\pm$1 sigma errors. The
bands intersect at the fit values for  $\phi_e$ and  $\phi_{\mu\tau}$, indicating
that the combined flux results are consistent with neutrino flavor transformation assuming no distortion in the $^8$B neutrino energy spectrum. From Q.~R.~Ahmad {\it et al.}  [SNO Collaboration],
Phys.\ Rev.\ Lett.\  {\bf 89}, 011301 (2002).
 \label{fig:SNO}}
\end{figure}

The ``bottom line'' of SNO are the comments written next to the detection reactions listed above. The three different reactions are sensitive to different neutrino flavors. Eq.~(\ref{SNO_CC}) is sensitive to only electron-type neutrinos, while Eq.~(\ref{SNO_ES}) is sensitive to mostly electron-type neutrinos, with a smaller cross-section for muon-type and tau-type neutrinos (0.15 is representative of the $\nu_{\mu,\tau}e$ to the $\nu_ee$ cross-section ration in the energy range of interest).
Eq.~(\ref{SNO_ES}) is the only reaction observed at the Super-Kamiokande detector. Finally, the neutral current process Eq.~(\ref{SNO_NC}) is flavor blind. In summary, SNO can measure, at the same time, the solar $\nu_e$ flux and the total ($\nu_{e}+\nu_{\mu}+\nu_{\tau}$) solar neutrino flux. Its results are summarized in Fig.~\ref{fig:SNO}. Its fair to say that SNO has established, beyond any reasonable doubt, that there are muon-type and/or tau-type neutrinos and/or antineutrinos coming from the Sun!

\section{Mass-Induced Flavor Change --- Neutrino Oscillations}
\label{oscillations}

To make a very long story short, it turns out that the only solution to all neutrino puzzles presented above is to appreciate that neutrinos can change flavor. Here, I'll discuss the only\footnote{I am aware of at least one other possible consistent explanation, (see, e.g., Ref.~\refcite{kostelecky_mewes}), which I'll not discuss. I'll just add that it (and others) has not been properly tested at the required level of precision.} scenario capable of fitting all data above in a unified and elegant way. This is related to postulating that neutrinos have mass.

Nonzero neutrino masses have been searched ever since the neutrino was first postulated (Pauli already quoted a very conservative upper bound of around 10~MeV in his 1930 letter). The most straight forward way to search for neutrino masses is to measure very precisely weak-processes that involve neutrinos in the final state. More specifically, the most stringent kinematical bounds on neutrinos masses are obtained by\cite{PDG}
\begin{itemize}
\item measuring the end-point of the $\beta$-decay spectrum of tritium: $m_{\nu_e}^2<5$~eV$^2$;
\item measuring the energy and momentum of the muon produced in pion decay at rest: $m_{\nu_{\mu}}<190$~keV;
\item measuring the total energy and momentum of pions produced by hadronic tau-decays $\tau\to N\pi+\nu_{\tau}$, $N>3$: $m_{\nu_{\tau}}<18.2$~MeV.
\end{itemize}
Above, $m_{\nu_{\alpha}}$ are linear combinations of the neutrino masses, as will be discussed later. Given what we have learned from the rest of the neutrino data, the bound on $m_{\nu_e}$ overwhelms the other two. 

Other more stringent and more subtle bounds can be obtained. Studies of the large-scale structure of the universe, combined with precision measurements of the cosmic microwave background radiation and the ``concordance cosmological model'' lead to a bound on the sum of all neutrino masses: $\sum m_{\nu}\lesssim 1$~eV.\cite{neutrino_cosmology} Finally, searches for neutrinoless double beta decay constrain a particular linear combination of neutrino masses --- $m_{\beta\beta}\lesssim 1$~eV --- assuming that neutrinos are Majorana fermions. I'll come back to this in Sec.~\ref{theend}.

\subsection{Lepton Mixing and Vacuum Oscillations}

The study of neutrino flavor change as a function of propagation distance (baseline) and neutrino energy turns out to be the most sensitive probe of neutrino masses, as long as lepton mixing is nontrivial.

Lepton mixing, like quark mixing,  can only be defined if the different neutrinos and charged leptons have distinct masses. There are different ways of understanding lepton mixing. One is to appreciate that, once neutrinos have distinct masses, there are two different ``types'' of neutrinos. There are the already-defined neutrino weak or flavor states. These are produced via charged current interactions and labeled by the charged lepton associated with the neutrino: $\nu_{\alpha}$ is produced/destroyed associated with the $\alpha$ charged lepton, where $\alpha=e,\mu,\tau$.   

On the other hand, there are the neutrino mass states. These are eigenstates of the free neutrino Hamiltonian, {\it i.e.},
\begin{equation}
|\nu_i(t,\vec{x})\rangle=e^{-ip_ix}|\nu_i(0,\vec{0})\rangle,~~~~p_i^2=m_i^2,
\end{equation}
where $i=1,2,3,\ldots$ labels the neutrino mass-eigenstates and mass-eigenvalues. It is clear that there is no reason for the neutrino mass-eigenstates to be the same as the neutrino flavor-eigenstates. The two different ``neutrino bases'' are related by a unitary transformation
\begin{equation}
|\nu_{\alpha}\rangle=U_{\alpha i}|\nu_i\rangle,
\label{UMNS}
\end{equation}
where $U_{\alpha i}$ are the elements of the neutrino, or lepton mixing matrix, also referred to as the Maki-Nakagawa-Sakata (MNS) matrix, or the Pontecorvo-Maki-Nakagawa-Sakata (PMNS, or MNSP) matrix. This means that, say, during $\beta$-decay, an electron and a linear combination of antineutrinos with well-defined masses are produced such that $m^2_{\nu_e}$ discussed above is given by\footnote{The dependency of the $\beta$-decay spectrum on the neutrino masses is a function of $m_{\nu_e}^2$ only in the limit where all neutrino masses are small enough.\cite{smirnov_xx}} $\sum_i |U_{ei}|^2m_i^2$.

Now, a more canonical description of fermion mixing. The relevant part of the weak-interaction Lagrangian is, assuming that the neutrinos are Dirac fermions and starting in the weak-basis where the charged-current interactions are diagonal,
\begin{eqnarray}
&{\rm L}\supset g\bar{e}_L^{\alpha} W_{\mu}\gamma^{\mu}\nu_L^{\alpha}+\bar{e}_L^{\alpha}m_{e,\alpha\beta}e_R^{\beta}+\bar{\nu}_L^{\alpha}m_{\nu,\alpha\beta}\nu_R^{\beta}+H.c. \nonumber \\
&=g\bar{e}_L^{\alpha} W_{\mu}\gamma^{\mu}\nu_L^{\alpha}+\bar{e}_L^{\alpha}(V_e^{\dagger})^{i\alpha}m^D_{e,ij}(U_e)^{j\beta}e_R^{\beta}+\bar{\nu}_L^{\alpha i}(V_{\nu}^{\dagger})^{\alpha i}m^D_{\nu,ij}(U_{\nu})^{j\beta}\nu_R^{\beta}+H.c. \nonumber \\
&=g\bar{e}^{\prime j}_L W_{\mu}\gamma^{\mu}(V_eV_{\nu}^{\dagger})^{ji}\nu^{\prime i}_L+m_{e,i}\bar{e}^{\prime i}_Le^{\prime i}_R+m_{\nu,i}\bar{\nu}_L^{\prime i}\nu_R^{\prime i}+H.c. 
\end{eqnarray}
where $V,U$ diagonalize the mass matrices, and relate the primed (mass) bases to the unprimed (weak) ones. The lepton analog of the CKM matrix is $U\equiv V_eV_{\nu}^{\dagger}$, and it is easy to show that it is identical to U defined by Eq.~(\ref{UMNS}).

Neutrinos are always produced and detected in well-defined flavor eigenstates. These, however, are not eigenstates of the propagation Hamiltonian. This mismatch leads to neutrino oscillations. As an example, assume that there are only two neutrino species, $\nu_{e}$ and $\nu_{\mu}$. An electron-type neutrino can be decomposed in terms of mass eigenstates $|\nu_1\rangle$ and $|\nu_2\rangle$ as
\begin{equation}
|\nu_e\rangle=\cos\theta|\nu_1\rangle+\sin\theta|\nu_2\rangle,
\end{equation}
where $\theta$ is the mixing angle that parameterizes the mixing matrix $U$.\footnote{A $2\times 2$ unitary matrix is parameterized by four real parameters. The other three parameters, however, turn out to be either unphysical or at least unobservable in the flavor oscillation phenomenon discussed here.} It is clear that the orthogonal muon-type neutrino state is $|\nu_{\mu}\rangle=-\sin\theta|\nu_1\rangle+\cos\theta|\nu_2\rangle$.

Assuming that the neutrino propagates as a plane-wave, at time $t$, the originally electron-neutrino state evolves into
\begin{equation}
|\nu(t,\vec{x})\rangle=\cos\theta e^{-ip_1x}|\nu_1\rangle+\sin\theta e^{-ip_2x}|\nu_2\rangle.
\end{equation}
The all-important phase factor is given by $p_ix=E_it-\vec{p}_i\vec{x}\simeq(E_i-p_{z,i})L$ ($i=1,2$) assuming that the neutrino is ultrarelativistic (always a very reasonable assumption) and travelling a distance $L$ along the $z$-direction. On the other hand, $E_i-p_{z,i}=(E_i^2-|\vec{p}|^2)/(E_i+p_{z,i})\simeq m_i^2/2E_i\simeq m_i^2/2E$ where $E_1\simeq E_2\simeq E$, and $E_i\simeq |\vec{p}_i|$. Hence
\begin{equation}
|\nu(L)\rangle=\cos\theta e^{-im_1^2L/2E}|\nu_1\rangle+\sin\theta e^{-im_2^2L/2E}|\nu_2\rangle.
\end{equation}
The probability that this state is an electron neutrino is
\begin{eqnarray}
&P_{ee}&=\left|\langle\nu_e|\nu(L)\rangle\right|^2, \nonumber \\ 
&&=\left|\left(\cos\theta\langle\nu_1|+\sin\theta\langle\nu_2|\right)
\left(\cos\theta e^{-im_1^2L/2E}|\nu_1\rangle+\sin\theta e^{-im_2^2L/2E}|\nu_2\rangle\right) 
\right|^2, \nonumber \\
&&=\left|\cos^2\theta e^{-im_1^2L/2E}+\sin^2\theta e^{-im_2^2L/2E} \right|^2, \nonumber \\
&&=\cos^4\theta+\sin^4\theta+2\sin^2\theta\cos^2\theta\Re\left(e^{-i(m_2^2-m_1^2)L/2E}\right), \nonumber \\
&&=1-4\cos^2\theta\sin^2\theta\left(\frac{1-\cos(\Delta m^2L/2E)}{2}\right), \nonumber \\
&&=1-\sin^22\theta\sin^2\left(\frac{\Delta m^2L}{4E}\right), \label{pee_2nus}
\end{eqnarray}
where $\Delta m^2\equiv m_2^2-m_1^2$ is the neutrino mass-squared difference. The unitary evolution of the neutrino state guarantees that $P_{ee}=P_{\mu\mu}=1-P_{e\mu}=1-P_{\mu e}$.

\subsubsection{Physics of Two-Flavor Vacuum Oscillations}

\begin{figure}[ht]
\centerline{\epsfig{width=0.8\textwidth, file=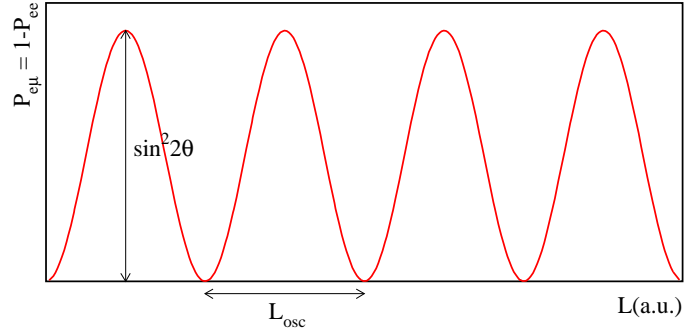}}
\caption{$P_{e\mu}$ in vacuum as a function of $L$ for fixed values of $E$, $\Delta m^2$, and $\sin^22\theta$. \label{osc_L}}
\end{figure}

Eq.~(\ref{pee_2nus}) dictates that an originally electron-type neutrino has a non-zero chance of being detected as a muon-type neutrino after it propagates a finite distance $L$. $P_{e\mu}$ as a function of $L$ for fixed $\Delta m^2$ and $E$ is depicted in Fig.~\ref{osc_L}. It is, of course, a periodic function of $L$. Its maximum is given by $\sin^22\theta$, and occurs every time $L=(2n+1)L_{\rm osc}/2$, $n=0,1,2,\ldots$, where $L_{\rm osc}$ is the neutrino oscillation length, defined as
\begin{equation}
\pi\frac{L}{L_{\rm osc}}\equiv \frac{\Delta m^2L}{4E}=1.267\left(\frac{L}{\rm km}\right)\left(\frac{\Delta m^2}{\rm eV^2}\right)\left(\frac{\rm GeV}{E}\right).
\end{equation}
Nontrivial effects are observed under two conditions. First, $\sin^22\theta$ should not be too small. Second, the neutrino oscillation length should not be much longer than the distance traversed by the neutrino. For particle physics-like neutrino energies (1~GeV), mass-squared differences of 1~eV$^2$ can be probed if the baseline is in the kilometer range.

It is useful to illustrate with a few examples. If neutrino oscillations in vacuum have anything to do with the solar neutrino puzzle ($E\sim 10$~MeV, $L=1$~astronomical unit) $\Delta m^2\sim 10^{-10}$ and $\sin^22\theta\sim 1$ is required. This possibility, referred to as the ``just-so'' solution, was ruled out by data from Super-Kamiokande and SNO. It is somewhat ironic that the oldest neutrino puzzle cannot be addressed by two flavor vacuum oscillations. It turns out that forward neutrino--electron scattering modifies the oscillation probabilities significantly, as will be discussed in the next subsection.

\begin{figure}[ht]
\centerline{\epsfig{width=0.8\textwidth, file=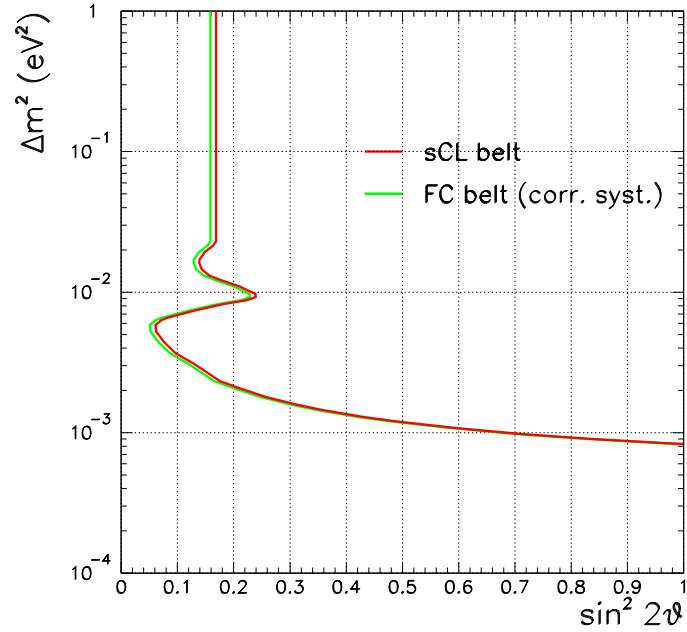}}
\caption{Exclusion plot at 90\% confidence level (sCL) for the oscillation parameters
based on the differential energy spectrum; the FC contour,
obtained with Òcorrect systematicsÓ treatment, is also shown. From M.~Apollonio {\it et al.},
Eur.\ Phys.\ J.\ C {\bf 27}, 331 (2003). \label{fig:chooz}}
\end{figure}

On the other hand, antineutrino fluxes from nuclear reactors have been measured far away from the reactor site. For example, the CHOOZ experiment in France,\cite{chooz} has measured the flux of electron-type reactor antineutrinos a little over one kilometer from the source. They have established that the observed flux agrees with the predicted one, and are able to set bounds on neutrino oscillation parameters, as depicted in Fig.~\ref{fig:chooz}. The shape of the exclusion curve is easy to understand. For $\Delta m^2L/4E\gg 1$, the oscillatory effects average out (remember that the reactor spectrum is continuos, and that the detector has a finite energy resolution) and $P_{ee}\simeq1-1/2\sin^22\theta$. The CHOOZ result, which can be translated into, roughly, $P_{ee}>0.95$, bounds $\sin^22\theta\lesssim 0.1$. On the other hand, in the limit $\Delta m^2L/4E\ll 1$, $P_{ee}\simeq 1-\sin^22\theta(\Delta m^2)^2(170)^2$ which leads to 
\begin{equation}
\left(\Delta m^2\right)^2\sin^22\theta\lesssim 10^{-6}.
\end{equation}
For many more details, please see Ref.~\refcite{chooz}.

The final example is related to the atmospheric neutrino data. It is fit very well by $\nu_{\mu}\leftrightarrow\nu_{\tau}$ oscillations with $\Delta m^2\sim 2\times 10^{-3}$~eV$^2$ (such that $L_{\rm osc}\sim 1000$~km for 1~GeV neutrinos) and $\sin^22\theta\sim 1$. For more details, see one of the homework problems in the appendix. The neutrino oscillation interpretation of the atmospheric data has, more recently, been confirmed by the K2K accelerator search for $\nu_{\mu}$ disappearance ($E\sim 1$~GeV, $L\sim 250$~km).\cite{K2K}

\subsection{Oscillations in the Presence of Matter}

The neutrino propagation equation, in the ultra-relativistic approximation, can be re-expressed in the form of a Shr\"odinger-like equation. Start with the mass-basis:
\begin{equation}
i\frac{\rm d}{{\rm d}L}|\nu_i\rangle=\frac{m_i^2}{2E}|\nu_i\rangle,
\end{equation}
up to a term proportional to the identity ($c|\nu_i\rangle$, where $c$ is a constant that does not depend on $i$), which is nonphysical (overall phase). 
By making use of $U^{\dagger}U=1$, and multiplying both sides of the equation above by $U_{\beta i}$
\begin{equation}
i\frac{\rm d}{{\rm d}L}|\nu_{\beta}\rangle=U_{\beta i}\frac{m_i^2}{2E}U^{\dagger}_{i\alpha}|\nu_{\alpha}\rangle.
\end{equation}
In the $2\times 2$ case, 
\begin{equation}
i\frac{\rm d}{{\rm d}L}\left( \begin{array}{c} |\nu_{e}\rangle \\ |\nu_{\mu}\rangle \end{array}\right)=\frac{\Delta m^2}{2E}\left(\begin{array}{cc} \sin^2\theta & \cos\theta\sin\theta \\ \cos\theta\sin\theta & \cos^2\theta \end{array} \right)\left(\begin{array}{c} |\nu_{e}\rangle \\ |\nu_{\mu}\rangle \end{array}\right),
\label{eq_osc_flavor}
\end{equation}
up to additional terms proportional to the $2\times 2$ identity matrix. Eq.~(\ref{eq_osc_flavor}) describes the propagation of flavor eigenstates, which contains, in general, non-diagonal terms and, hence, mixing.

Let us re-examine the Fermi Lagragian, concentrating on the electron-type neutrinos and their interaction with electrons. After a Fiertz rearrangement of the charged-current terms, the Lagrangian is
\begin{equation}
{\rm L}\supset\bar{\nu}_{eL}i\partial_{\mu}\gamma^{\mu}\nu_{eL}-2\sqrt{2}G_F\left(\bar{\nu}_{eL}\gamma^{\mu}\nu_{eL}\right)\left(\bar{e}_{L}\gamma_{\mu}e_{L}\right)+\ldots.
\end{equation}
Given the Lagrangian above, we wish to compute the equation of motion for one electron neutrino state in the presence of a non-relativistic electron background. In this case, we need to compute
\begin{equation}
\langle\bar{e}_L\gamma_{\mu}e_{L}\rangle=\delta_{\mu0}\frac{N_e}{2}
\end{equation}
where $N_e\equiv e^{\dagger}e$ is the average electron number density (which is at rest, hence the $\delta_{\mu0}$ part), and the factor of 1/2 comes from the fact that half of the electron number density is right-handed, while the other half is left-handed. The neutrinos only see the left-handed half.

Ignoring mass-terms for the time being, the Dirac equation for a one neutrino state inside a cold electron ``gas'' is
\begin{equation}
(i\partial^{\mu}\gamma_{\mu}-\sqrt{2}G_FN_e\gamma_0)|\nu_e\rangle=0.
\label{Dirac_matter}
\end{equation}
Note that Eq.~(\ref{Dirac_matter}) is not Lorentz invariant. Its solutions are still plane-wave like and, in the ultrarelativistic limit and in the limit that $\sqrt{2}G_FN_e\ll E$, the neutrino dispersion relation is
\begin{equation}
E\simeq \left|\vec{p}\right|\pm \sqrt{2}G_FN_e,
\end{equation}
where the plus sign applies to the positive energy solutions (neutrinos) and the minus one to the negative energy ones (antineutrinos). The modified dispersion relation of neutrinos propagating in matter is similar to the modified dispersion relation of photons propagating inside matter (index of refraction).

$\sqrt{2}G_FN_e$ is referred to as the matter potential, because it looks like a ``potential energy'' term for the neutrino (using a classical mechanics analogy, $E=T+V$).

It is easy to see how the effects of matter will change Eq.~(\ref{eq_osc_flavor}):
\begin{equation}
i\frac{\rm d}{{\rm d}L}\left( \begin{array}{c} |\nu_{e}\rangle \\ |\nu_{\mu}\rangle \end{array}\right)=\left[\frac{\Delta m^2}{2E}\left(\begin{array}{cc} \sin^2\theta & \cos\theta\sin\theta \\ \cos\theta\sin\theta & \cos^2\theta \end{array} \right)
+\left(\begin{array}{cc}A & 0 \\ 0 & 0\end{array}\right)
\right]\left(\begin{array}{c} |\nu_{e}\rangle \\ |\nu_{\mu}\rangle \end{array}\right),
\label{eq_osc_matter}
\end{equation}
where $A=\pm\sqrt{2}G_FN_e$ ($+$ for neutrinos, $-$ for antineutrinos). A similar effect also comes from neutral current interactions. These, however, are common to all (active) neutrino species, and translate into a term in Eq.~(\ref{eq_osc_matter}) proportional to the identity matrix.

Eq.~(\ref{eq_osc_matter}) is not easy to solve in general. $A$ is proportional to the electron number density along the path of the neutrino, which can be a complicated function of $L$. Eq.~(\ref{eq_osc_matter}) can be thought of as a two-level non-relativistc quantum mechanical system in the presence of an external potential which can be ``time'' dependent.

Under several conditions, however, Eq.~(\ref{eq_osc_matter}) can be solved exactly, and I'll discuss two of the most useful ones. The first obvious approximation is to assume that $A$ is a constant. This is a very good approximation for neutrinos propagating through matter inside the Earth, with the exception of neutrinos that traverse different ``Earth layers'' (the crust, the mantle, the outer core, the inner core). 

Rewrite Eq.~(\ref{eq_osc_matter}) as
\begin{equation}
i\frac{\rm d}{{\rm d}L}\left( \begin{array}{c} |\nu_{e}\rangle \\ |\nu_{\mu}\rangle \end{array}\right)=\left(\begin{array}{cc} A & \Delta/2\sin2\theta \\ \Delta/2\sin2\theta & \Delta\cos2\theta \end{array} \right)\left(\begin{array}{c} |\nu_{e}\rangle \\ |\nu_{\mu}\rangle \end{array}\right),
\label{eq_osc_constant}
\end{equation}
where $\Delta\equiv\Delta m^2/2E$. By comparing Eq.~(\ref{eq_osc_constant}) to Eq.~(\ref{eq_osc_flavor}), it is easy to guess that ({\it cf.} Eq.~(\ref{pee_2nus}))
\begin{equation}
P_{e\mu}=\sin^22\theta_M\sin^2\left(\frac{\Delta_ML}{2}\right),
\label{pemu_matter}
\end{equation}
where $\theta_M$ is some effective matter mixing angle characterisitic of the eigenvectors of the Hamiltonian in Eq.~(\ref{eq_osc_constant}), while $\Delta_M$ is the difference between the eigenvalues of the Hamiltonian in Eq.~(\ref{eq_osc_constant}). $\Delta_M$ is also proportional to the effective inverse oscillation length in matter. 
Some trivial linear algebra reveals
\begin{eqnarray}
\Delta_M&=&\sqrt{\left(A-\Delta\cos2\theta\right)^2+\Delta^2\sin^22\theta}, \label{deltaM} \\
\Delta_M\sin2\theta_M&=&\Delta \sin2\theta, \label{sintM}\\
\Delta_M\cos2\theta_M&=&A-\Delta\cos2\theta.
\end{eqnarray}
The presence of matter affects neutrino and antineutrino oscillation differently. This effective ``CPT-violation'' is expected, given that we are assuming that the neutrinos are propagating in a background of {\sl electrons}. The CPT-theorem relates the propagation of neutrinos in an electron background to the propagation of antineutrinos in a positron background.

Furthermore, the presence of matter allows one to explore the neutrino mass and mixing landscape ``more.'' It is instructive to ask what is the ``physical parameter'' space of two-flavor neutrino oscillations, {\it i.e.}, what are the values of $\theta$ and $\Delta m^2$ that span all the distinct physical circumstances? Here, I'll choose $m_2^2\ge m_1^2$ (one can view this as the definition of $\nu_1$ and $\nu_2$ as, respectively, the lighter and the heavier neutrino), such that $\Delta m^2\in [0,\infty\}$. Under these circumstances, $\theta=0$ corresponds to $|\nu_e\rangle=|\nu_1\rangle$ (the lighter state), while $\theta=\pi/2$ corresponds to $|\nu_e\rangle=|\nu_2\rangle$ (the heavier state). Given that oscillation experiments are not sensitive to the relative phase between the $|\nu_1\rangle$ and $|\nu_2\rangle$ components of $|\nu_e\rangle$,\footnote{Check this (or see, for example, Ref.~\refcite{darkside})!} $\theta\in[0,\pi/2]$ describes all physically distinguishable circumstances. Eq.~(\ref{pee_2nus}), however, is invariant under $\theta\leftrightarrow\pi/2-\theta$, such that it cannot tell whether the $|\nu_e\rangle$ state is ``mostly light'' ($\cos^2\theta>\sin^2\theta$, the ``light-side''\cite{darkside}) or ``mostly heavy'' ($\cos^2\theta<\sin^2\theta$, the ``dark-side''\cite{darkside}).

\begin{figure}[ht]
\centerline{\epsfig{width=0.8\textwidth, file=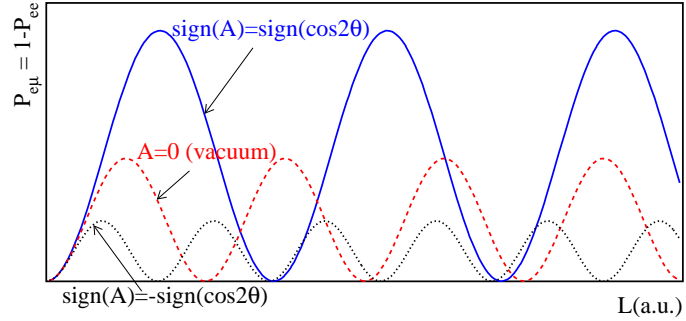}}
\caption{$P_{e\mu}$ as a function of $L$ for fixed values of $E$, $\Delta m^2$, $A$, and $\sin^22\theta$ in vacuum and in matter, assuming sign$(A)=$sign$(\cos2\theta)$ and   sign$(A)=-$sign$(\cos2\theta)$. \label{fig:osc_matter}}
\end{figure}

Oscillations in matter, however, do not suffer from the same degeneracy problem. Since Eq.~(\ref{deltaM}) depends on $\cos2\theta$, oscillations in matter are sensitive to the entire two-flavor oscillation parameter space. Fig.~\ref{fig:osc_matter} depicts $P_{e\mu}$ in vacuum and in matter, assuming that the sign of $A$ agrees or disagrees with the sign of $\cos2\theta$. It is clear that when the two signs agree (disagree), there is an enhancement (supression) of the transition amplitude, $\sin^22\theta_M> (<)\sin^22\theta$. Optimal enhancement can be obtained when $A=\Delta\cos2\theta$, the so-called resonant condition, in which case $\sin^22\theta_M=1$. It is also clear that the oscillation length increases (decreases) with respect to the vacuum one in the case of a matter enhancement (suppression). Another interesting feature is the fact that, for small $L$, matter effects ``don't matter.'' This is very easy to see by plugging Eq.~(\ref{sintM}) into Eq.~(\ref{pemu_matter}) in the limit $\Delta_ML,\Delta L\ll1$.

\subsubsection{Varying Electron Number Density --- The MSW Effect}

It is curious that in order to understand the oldest neutrino puzzle, one is required to use more advanced ``technology'' than the one developed above.\cite{MSW} Solar neutrinos are created deep inside the Sun where the matter density is very high and propagate outward until they eventually meet ``empty space.'' Along the way, the electron number density varies, to a reasonably good approximation, exponentially.\cite{bahcall_web} 

In general, there is no exact solution to the propagation of neutrinos in matter of varying density. However, if certain approximations apply, a nice qualitative understanding can be obtained.\cite{kuo_pantaleone} This is, fortunately, the case for solar neutrinos and neutrinos from other astrophysical sources.

\begin{figure}[ht]
\centerline{\epsfig{width=0.7\textwidth, file=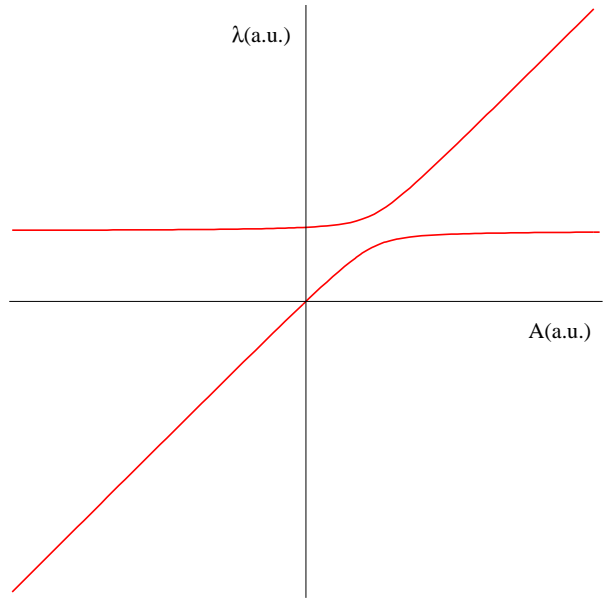}}
\caption{Eigenstates $\lambda$ of the Hamiltonian Eq.~(\ref{Ham}) as a function of $A$, for fixed $\Delta$ and $\theta$, modulo a constant term, common to both eigenstates. \label{levelcrossing}}
\end{figure}

First, consider the ``Hamiltonian'' of Eq.~(\ref{eq_osc_matter}), which I reproduce again below
\begin{equation}
\left[\Delta\left(\begin{array}{cc} \sin^2\theta & \cos\theta\sin\theta \\ \cos\theta\sin\theta & \cos^2\theta \end{array} \right)
+A\left(\begin{array}{cc}1 & 0 \\ 0 & 0\end{array}\right)
\right],
\label{Ham}
\end{equation}
and compute its eigenvalues as a function of $A$ (for fixed $\Delta$ and $\theta$). These are depicted in Fig.~\ref{levelcrossing} in the case $\cos2\theta>0$.

Assume that one starts at very large, positive, values of $A$ (say, inside the Sun) such that $|\nu_e\rangle\simeq|\nu_{2M}\rangle$, the heavier Hamiltonian eigenstate associated to this value of $A$ (this is easy to see from Eq.~(\ref{Ham}), in the limit $A\gg\Delta$). Let us further assume that $A$ decreases ``slowly'' as a function of $L$, such that the system evolves adiabatically. In this case, the initial $|\nu_e\rangle$ will track the heavy ``instantaneous Hamiltionian eigenstate,'' such that, when $A$ reaches zero (say, at the ``end'' of the Sun), the original $|\nu_e\rangle$ is now a $|\nu_2\rangle$ state. In summary
\begin{eqnarray}
&|\nu_e\rangle=|\nu_{2M}\rangle~{\rm at~the~core}~\to |\nu_2\rangle~{\rm in~vacuum}, \\
 &P_{ee}^{\rm Earth}=\left|\langle\nu_e|\nu_2\rangle\right|^2=\sin^2\theta.
\end{eqnarray}
Note that $P_{ee}\simeq \sin^2\theta$ applies in a wide range of energies and baselines, as long as the  approximations mentioned above apply --- ideal to explain the energy independent suppression of the $^8$B solar neutrino flux! Furthermore, $\sin^2\theta\in[0,1]$, which means that large average suppressions of the neutrino flux are allowed if $\sin^2\theta\ll1$. This is to be compared with averaged out vacuum oscillations $\bar{P}_{ee}^{\rm vac}=1-1/2\sin^22\theta>1/2$. 

One can expand on the result above by loosening some of the assumptions. For example, assume that a $|\nu_e\rangle$ state is produced in the Sun's core as an {\sl incoherent} mixture of $|\nu_{1M}\rangle$ and $|\nu_{2M}\rangle$.\footnote{Numerically, this is an excellent approximation, due to the fact that the neutrino production region inside the Sun is ``large,'' while the matter oscillation length inside of the Sun is ``small.''} Furthermore, introduce an adiabaticity parameter $P_c$, which measures the probability that a $|\nu_{iM}\rangle$ matter Hamiltonian state will {\sl not} exit the Sun as a $|\nu_i\rangle$ mass-eigenstate. Hence
\begin{eqnarray}
|\nu_e\rangle&\to&|\nu_{1M}\rangle,~{\rm with~probability}~\cos^2\theta_M, \\
&\to&|\nu_{2M}\rangle,~{\rm with~probability}~\sin^2\theta_M,
\end{eqnarray}
where $\theta_M$ is the matter angle at the neutrino production point. Furthermore,
\begin{eqnarray}
|\nu_{1M}\rangle&\to&|\nu_{1}\rangle,~{\rm with~probability}~(1-P_c), \\
&\to&|\nu_{2}\rangle,~{\rm with~probability}~P_c, \\
|\nu_{2M}\rangle&\to&|\nu_{1}\rangle~{\rm with~probability}~P_c,  \\
&\to&|\nu_{2}\rangle~{\rm with~probability}~(1-P_c). 
\end{eqnarray}
Finally, since $P_{1e}=\cos^2\theta$ and $P_{2e}=\sin^2\theta$, one can simply read off
\begin{eqnarray}
P_{ee}^{\rm Sun}=&\cos^2\theta_{M}\left[(1-P_c)\cos^2\theta+P_c\sin^2\theta\right] \nonumber \\
&+\sin^2\theta_M\left[P_c\cos^2\theta+(1-P_c)\sin^2\theta\right].
\label{pee_sun}
\end{eqnarray}
For an exponential electron number density $N_e=N_{e0}e^{-L/r_0}$, $P_c$, also referred to as the crossing probability, is exactly calculable\cite{pc_exp}
\begin{equation}
P_c=\frac{e^{-\gamma\sin^2\theta}-e^{-\gamma}}{1-e^{-\gamma}},~~\gamma=2\pi r_0\Delta.
\end{equation}
Note that for $\gamma\gg1$, $P_c\to 0$ as expected, since $\gamma\gg1$ implies $1/r_0\ll\Delta$, {\it i.e.}, the natural frequency of the system is much larger than the rate of change of the potential, which is the adiabatic condition.

\begin{figure}[ht]
\centerline{\epsfig{width=0.7\textwidth, file=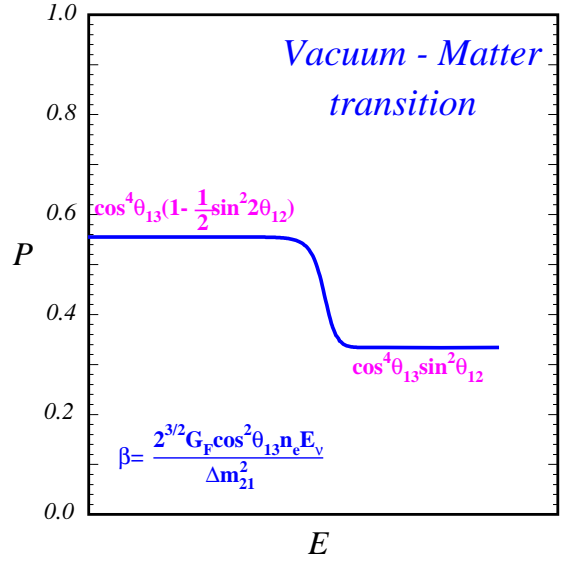}}
\caption{Electron neutrino survival probability, $P_{ee}$, as a function of neutrino energy for the (daytime) LMA oscillation solution.  For small values of the parameter $\beta$, the vacuum (kinematic) oscillation effects are dominant.  For values of $\beta$ greater than unity, the MSW (matter) oscillations are most important.  In order to properly fit the solar neutrino data, the transition between vacuum and matter oscillations should occur somewhere in the region of 2 MeV. In the case of two flavor oscillations, set $\theta_{13}=0$, and $\theta_{12}=\theta$. From http://www.sns.ias.edu/$\sim$jnb and J.N.~Bahcall and C.~Pe\~na-Garay, JHEP {\bf 0311}, 004 (2003). \label{sol_survival}}
\end{figure}

The best application of Eq.~(\ref{pee_sun}) is in solving the solar neutrino puzzle. In particular, for $P_c=0$, the qualitative description of $P_{ee}$ is depicted in Fig.~\ref{sol_survival}. This particular shape is ideal to fit the solar data as long as one requires $\sin^2\theta\sim 0.3$ and $\Delta m^2\sim 10^{-5}$~eV$^2$ to $10^{-4}$~eV$^2$.

\begin{figure}[ht]
\centerline{\epsfig{width=1\textwidth, file=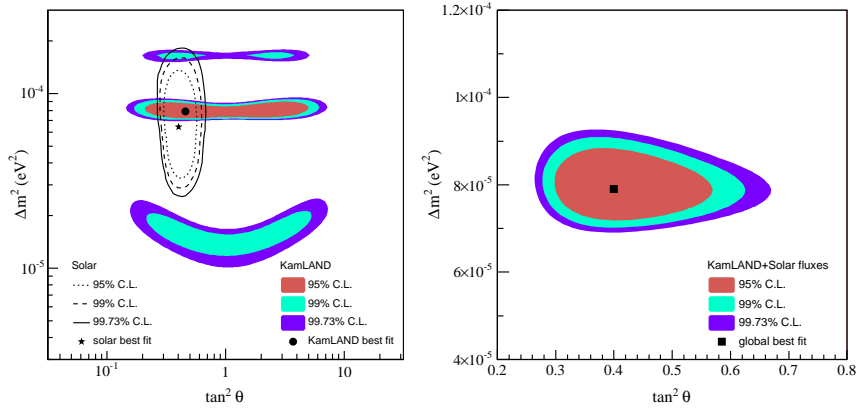}}
\caption{LEFT -- Neutrino oscillation parameter allowed region from
  KamLAND anti-neutrino data (shaded regions) and solar neutrino
  experiments (lines). RIGHT -- Result of a combined
  two-neutrino oscillation analysis of KamLAND and the observed solar
  neutrino fluxes. The fit
  gives $\Delta m^2=7.9^{+0.6}_{-0.5}\times 10^{-5}$~eV$^{2}$ and
  $\tan^2\theta=0.40^{+0.10}_{-0.07}$ including the allowed one sigma
  parameter range. From T.~Araki {\it et al.}  [KamLAND Collaboration], hep-ex/0406035.
\label{kam_res}}
\end{figure}

The solar neutrino puzzle was ultimately resolved by the KamLAND reactor antineutrino experiment. For $\Delta m^2$ values in the preferred solar range, reactor antineutrinos, with typical energies in the 1--10~MeV range, have an oscillation length $L_{\rm osc}\lesssim 2\times 10^{5}$~m. Hence, if one could measure the reactor antineutrino flux some 100~km away from a nuclear reactor, one should observe an order one suppression (because the solar data implies a large mixing angle) of the antineutrino flux. The catch is that the expected flux from a nuclear reactor 100~km away is tiny (remember that $\Phi\propto L^{-2}$ in the universe we live in, with 3+1 large spacetime dimensions), which requires a very big detector or a really powerful nuclear power plant.  The KamLAND collaboration addressed this problem by building a very large liquid scintillator detector (around 1~kton) and placing it around 100~km away from {\sl several} nuclear reactors in Japan. It is a very impressive achievement of our understanding of neutrino and solar physics that the results obtained by KamLAND regarding the neutrino oscillation parameters agree perfectly with the results of the solar experiments. This agreement in depicted in Fig.~\ref{kam_res}.

\section{Solving the Neutrino Puzzles}
\label{solve_puzzle}

In order to fit solar, atmospheric, reactor and accelerator neutrino data (except for the LSND anomaly, {\it cf.} Sec.~\ref{sec_lsnd}), it is clear that two-flavor oscillations do not suffice. The reason for this is obvious: among other things, solar/KamLAND and atmospheric/K2K data point to very different values of the mass-squared difference and the mixing angle. Of course, the situation is naively remedied once one remembers that we are aware of three  neutrinos --- not two. 


For three neutrino flavors, the MNS matrix is defined as
\begin{equation}
\left(\begin{array}{c}\nu_e \\ \nu_{\mu} \\ \nu_{\tau} \end{array} \right) =\left(\begin{array}{ccc} U_{e1} & U_{e2} & U_{e3} \\ U_{\mu1} & U_{\mu2} & U_{\mu3} \\ U_{\tau1} & U_{e\tau2} & U_{\tau3}\end{array}\right)
\left(\begin{array}{c}\nu_1 \\ \nu_2 \\ \nu_3 \end{array} \right),
\label{UMNS3}
\end{equation}
and its elements are, of course, not all independent. It is customary\cite{PDG} to parameterize $U$ in Eq.~(\ref{UMNS3}) with three mixing angles $\theta_{12},\theta_{13},\theta_{23}$ and three complex phases, $\delta,\xi,\eta$, defined by
\begin{equation}
\frac{|U_{e2}|^2}{|U_{e1}|^2}\equiv \tan^2\theta_{12};
~~~~\frac{|U_{\mu3}|^2}{|U_{\tau3}|^2}\equiv \tan^2\theta_{23};~~~~
U_{e3}\equiv\sin\theta_{13}e^{-i\delta},
\end{equation}
with the exception of $\xi$ and $\eta$, the so-called Majorana CP-odd phases. These are only physical if the neutrinos are Majorana fermions, and have, unfortunately, virtually no effect in flavor-changing phenomena.\footnote{for a more detailed discussion of Majorana phases and their physical effects, see, for example, Ref.~\refcite{cp_maj}} We have no idea what their values are or even whether they are physical observables!

\begin{figure}[ht]
\centerline{\epsfig{width=0.6\textwidth, file=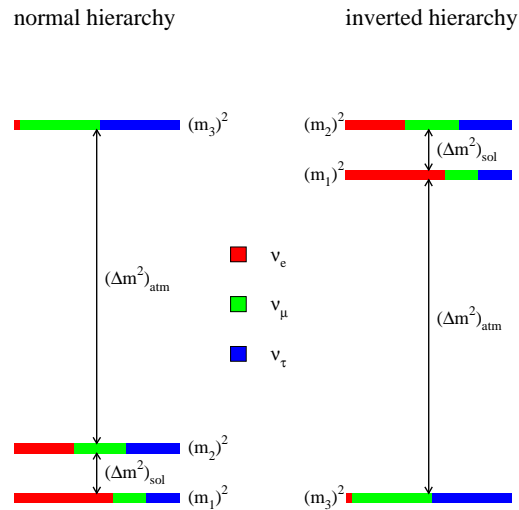}}
\caption{Cartoon of the two distinct neutrino-mass hierarchies that fit all of the current neutrino data, for fixed values of all mixing angles and mass-squared differences. The color coding (shading) indices the fraction $|U_{\alpha i}|^2$ of each distinct flavor $\nu_{\alpha}$, $\alpha=e,\mu,\tau$ contained in each mass eigenstate $\nu_i$, $i=1,2,3$. For example, $|U_{e2}|^2$ is equal to the fraction of the $(m_2)^2$ ``bar'' that is painted red (shading labeled as `$\nu_e$').
\label{normal_inverted}}
\end{figure}

In order to proceed unambiguously, it is necessary to define the neutrino mass eigenstates, {\it i.e.}, to ``order'' the neutrino masses. This is most often done in the following way: $m_2^2>m_1^2$ and $\Delta m^2_{12}<|\Delta m^2_{13}|$. In this case, there are three mass-related observables: $\Delta m^2_{12}$ (positive definite), $|\Delta m^2_{13}|$, and the sign of $\Delta m^2_{13}$. A positive sign for $\Delta m^2_{13}$ implies $m_3^2>m_2^2$ --- a so-called normal mass-hierarchy --- while a negative sign for $\Delta m^2_{13}$ implies $m_3^2<m_1^2$  --- a so-called inverted mass-hierarchy. The two distinct neutrino mass-hierarchies are depicted in Fig.~\ref{normal_inverted}. Finally, the data tell us that the ratio $\Delta m^2_{12}/|\Delta m^2_{13}|\lesssim 1/30$ is small. We will take full advantage of that in what follows.

In vacuum, the oscillation probabilities can be written as
\begin{eqnarray}
P_{\alpha\beta}&=&\delta_{\alpha\beta}+A^{\rm sol}_{\alpha\beta}\sin^2\left(\frac{\Delta m^2_{12}L}{4E}\right)+A^{\rm atm}_{\alpha\beta}\sin^2\left(\frac{\Delta m^2_{13}L}{4E}\right)+ \nonumber \\ 
&&+F_{\alpha\beta}^{\rm int}(L,\Delta m^2_{12},\Delta m^2_{13}),
\end{eqnarray}
where $A_{\alpha\beta}$ are the ``solar'' and ``atmospheric'' amplitudes, functions of $U_{\alpha i}$, while $F_{\alpha\beta}$ are ``interference terms.''\footnote{I'll leave as an exercise to show that this is the case, and to figure out what the expressions for the $A$'s and $F$'s are.} 

The survival probability of atmospheric muon-type neutrinos is given by
\begin{equation}
P_{\mu\mu}^{\rm atm}\simeq 1-4|U_{\mu3}|^2\left(1-|U_{\mu3}|^2\right)\sin^2\left(\frac{\Delta m^2_{13}L}{4E}\right),
\label{atm3}
\end{equation}
ignoring effects from the ``solar'' oscillation length, which will turn out to be much longer than the the Earth's diameter for typical atmospheric  neutrino energies. For the same reason, the survival probability of reactor electron-type antineutrinos at the CHOOZ experiment ($L\sim 1$~km) can be written as
\begin{equation}
P_{ee}^{\rm CHOOZ}\simeq 1-4|U_{e3}|^2\left(1-|U_{e3}|^2\right)\sin^2\left(\frac{\Delta m^2_{13}L}{4E}\right).
\label{chooz3}
\end{equation}
Eq.~(\ref{atm3}) and Eq.~(\ref{chooz3}) are effective two flavor oscillation expressions, and the data require (i) $\Delta m^2_{13}\simeq 2\times 10^{-3}$~eV$^2$, (ii) $|U_{\mu3}|^2\simeq 0.5$, and (iii) $|U_{e3}|^2\equiv\sin^2\theta_{13}\lesssim0.05$. Unitarity of $U$ combined with (ii) and (iii) lead to $|U_{\tau3}|^2\simeq 0.5$, such that $\tan^2\theta_{23}\simeq 1$.

The survival probability of electron-type neutrinos from the Sun and antineutrinos from reactors at KamLAND is given by
\begin{equation}
P_{ee}\simeq \cos^4\theta_{13}P_{ee}^{2\nu}(\Delta m^2_{12},\sin^2\theta_{12})+\sin^4\theta_{13},
\label{pee3}
\end{equation} 
where $P_{ee}^{2\nu}$ is the appropriate two-flavor neutrino oscillation equation with $\Delta m^2=\Delta m^2_{12}$, $\sin^2\theta=\sin^2\theta_{12}$, and a modified matter potential $A\to A\cos^2\theta_{13}$. The reasons behind the validity of  Eq.~(\ref{pee3}) are (i) ``atmospheric'' effects average out given that the associated oscillation lengths are much smaller than the baselines of the KamLAND experiment and (ii) the matter potential in the Sun's core is significantly smaller than $|\Delta m^2_{13}|/(2E)$ for solar neutrino energies. 

Again, Eq.~(\ref{pee3}) is almost an effective two-flavor oscillation expression (especially because $\theta_{13}$ is known to be small), and the data point to $\Delta m^2_{12}\sim 10^{-4}$~eV$^2$ and $\sin^2\theta_{12}\sim 0.3$.

\begin{figure}[ht]
\centerline{\epsfig{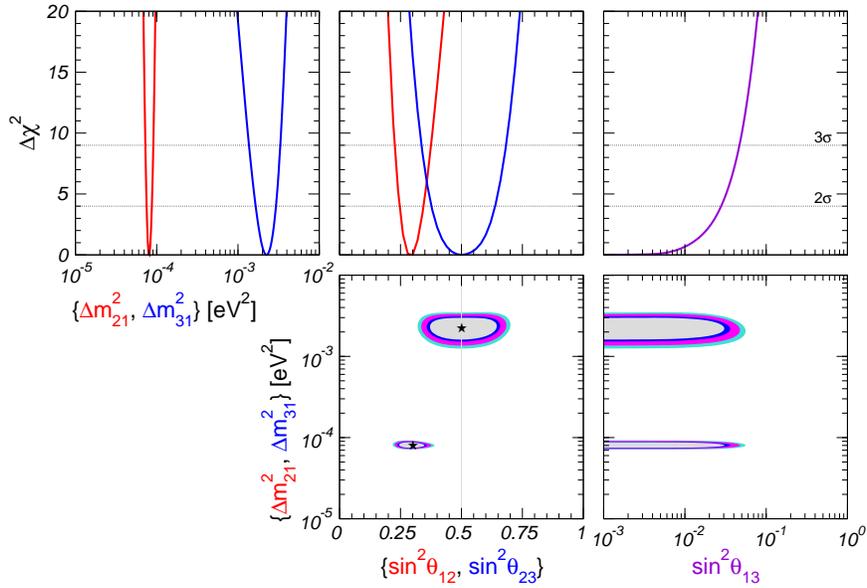}}
\caption{Projections of the allowed regions from the global oscillation
      data at 90\%, 95\%, 99\%, and 3$\sigma$ confidence levels for two degrees of freedom for
      various parameter combinations. Also shown are the values of $\Delta \chi^2$ as
      a function of the oscillation parameters $\sin^2\theta_{12},
      \sin^2\theta_{23}, \sin^2\theta_{13}, \Delta m^2_{12}, \Delta
      m^2_{13}$, minimized with respect to all undisplayed
      parameters. From M.~Maltoni, T.~Schwetz, M.A.~Tortola and J.W.F.~Valle, hep-ph/0405172. \label{full_fit}}
\end{figure}

Detailed combined analyses of all neutrino data are consistent, at the three sigma confidence level, with\cite{valle_new}
\begin{itemize}
\item $\sin^2\theta_{12}=0.30\pm0.08$, mostly from solar data;
\item $\sin^2\theta_{23}=0.50\pm0.18$, mostly from atmospheric neutrino data;
\item $\sin^2\theta_{13}\leq 0047$, mostly from atmospheric and CHOOZ data;
\item $\Delta m^2_{12}=8.1\pm1.0$~eV$^2$, mostly from KamLAND data;
\item $|\Delta m^2_{13}|=2.2\pm1.1$~eV$^2$, mostly from atmospheric neutrino data.
\end{itemize}
These results are summarized in Fig.~\ref{full_fit}.\footnote{In the figure, $\Delta m^2_{ij}$ is defined by $m_i^2-m_j^2$, as opposed to the one I adopt here, $m_j^2-m_i^2$.} As one can clearly see, in the span of less than a decade we have ``evolved'' from not being sure whether neutrinos had mass to ``precisely'' measuring the neutrino mass-squared differences and the elements of the leptonic mixing matrix. But a lot of work is still left for the next-generation of neutrino experiments (and ``neutrino theorists!'').

\subsection{Plans for the Future}

As far as neutrino oscillation experiments are concerned, and assuming there are no other surprises lurking somewhere, we still need to find out 
\begin{itemize}
\item the sign of $\Delta m^2_{13}$ or what is the neutrino mass hierarchy?;
\item the value of $\theta_{13}$ or is $|U_{e3}|\neq 0$?;
\item the value of $\delta$ or is there CP-invariance violation in neutrino oscillations?;
\item the value of $1/2-\sin^2\theta_{23}$ or how close to maximal is $\nu_{\mu}\leftrightarrow\nu_{\tau}$ mixing?
\end{itemize}
Furthermore, we need to know several of the already-measured oscillation parameters more precisely in order to start filling in the gaps. Current, near-future, and far-future experiments are being built/planned/considered in order to fully piece together the neutrino flavor-change phenomenon.

There are ``on-going'' long baseline experiments. All of these consist of a muon-neutrino beam, just like the one used to discover the muon-type neutrino, aimed at a detector some hundreds of kilometers away. These include
\begin{itemize}
\item The K2K experiment (``KEK to Kamionka''), already alluded to and currently taking data in Japan. K2K studies the disappearance of muon-type neutrinos in the ``atmospheric frequency.'' K2K data agree with Super-Kamiokande atmospheric neutrino data, and the collaboration has recently reported a hint of an ``oscillatory suppression'' of the $\nu_{\mu}$-flux;\cite{K2K}
\item The MINOS experiment (``Main Injector Neutrino Oscillation Search''), is being built at Fermilab and should start data taking in the end of 2004.\cite{minos} It is similar to the K2K experiment, with a longer baseline ($L=732$~km) and higher neutrino energies ($E\sim 1-10$~GeV). Minos has the ability to see a more pronounced oscillatory behavior of the muon spectrum at the far detector, and will be able to measure $\Delta m^2_{13}$ more precisely than K2K. It is also more sensitive to $\nu_{\mu}\to\nu_e$ appearance driven by a nonzero $|U_{e3}|^2$ than CHOOZ was sensitive to $\bar{\nu}_e$ disappearance (by about a factor of two).
\item The CNGS (``CERN Neutrinos to Gran Sasso'') project is being constructed at CERN, and aims at starting data taking in a few years.\cite{cngs} Its baseline is similar to the MINOS one, but the typical neutrino energies are much higher (tens of GeV). Among other things, it is sensitive to tau-appearance driven by the expected ``atmospheric'' $\nu_{\mu}\leftrightarrow\nu_{\tau}$ oscillations. 
\end{itemize}

Near-future oscillation experiments are either under construction or their proposals are (in most cases) under review.\footnote{A recent study of the future of neutrino physics has recently become available.\cite{APS} It describes future experiments in much more detail, and contain a large amount of information.}  I'll highlight two classes of experiments:
\begin{itemize}
\item Long baseline experiments optimized to study $\nu_{\mu}\to\nu_e$ transitions governed by the atmospheric mass-squared difference. In this case 
\begin{equation}
P_{e\mu}\propto |U^{\rm matter}_{e3}|^2\cos^2\theta_{23}\sin^2\left(\frac{\Delta^{\rm matter}_{13}L}{2}\right)+\ldots
\end{equation}
where the `matter' superscripts indicate that matter effects may be sizeable, and the ellipsis indicate sub-leading effects due to ``solar'' oscillations.

These setups are not only sensitive to $|U_{e3}|^2$, but are also sensitive to the sign of $\Delta m^2_{13}$, as long as the matter effects are significant. Furthermore, they are also sensitive to effects of the CP-odd phase $\delta$. 

\item Next-generation reactor experiments, with baselines around one kilometer. Here the oscillation probability is given by Eq.~(\ref{chooz3}), and the goal is to improve on the sensitivity of the CHOOZ experiment by an order of magnitude. This is hoped to be achieved by optimizing the reactor--detector distance, and adding a near-detector, capable of measuring the reactor antineutrino flux and reducing systematic errors significantly.\cite{reactor_white} Complementary to long-baseline $\nu_{\mu}\to\nu_e$ studies, these setups are only sensitive to $|U_{e3}|^2$. Hence, they can obtain a very ``clean'' measurement of $|U_{e3}|^2$, but do not have the ability to determine the mass-hierarchy or whether CP-invariance is violated in the neutrino oscillations.  
\end{itemize}

Further in the future, the neutrino community is contemplating building completely new neutrino facilities, capable of providing different, precisely calculable neutrino beams. Among them are
\begin{itemize}
\item Neutrino Factories, or muon storage rings. The idea is to accelerate and ``store'' muons in a ring-like structure.\cite{nufact} Natural muon decay produces an intense, very well known beam of muon-type neutrinos and electron-type antineutrinos (or vice-versa, depending on the muon-charge). Note that neutrino factories offer a very intense, {\sl high energy} electron-type neutrino beam, something we have not been able to work with so far! This allows one to study $\nu_e\to\nu_{\mu}$ transitions (as opposed to $\nu_{\mu}\to\nu_{e}$). $\nu_e\to\nu_{\mu}$  not only serves as a new channel to probe neutrino flavor change, but it is often the case that high-energy muons are easier to study in typical neutrino detectors than electrons.  

\item More recently, the community has been considering a different option for generating a high energy electron-type neutrino beam --- $\beta-$beams.\cite{betabeam} These consist of accelerating to large $\gamma$-factors and storing $\beta^-$ and $\beta^+$ decaying nuclei (say, $^6$He$\to^6$Li$+e^-+\bar{\nu}_e$ and $^{18}$Ne$\to^{18}$F$+e^++\nu_e$) and using them as sources of electron-type antineutrinos or neutrinos, respectively. 
\end{itemize}

Needless to say, there are many physics and technology (and financial) issues that need to be resolved before either neutrino factories or beta-beam facilities can be built. They are, nonetheless, the subject of serious research and development activity and are viewed as the most powerful tools to study neutrinos in particular, and lepton physics in general. They may ultimately be required in order for us to obtain a deeper and more satisfying understanding of neutrinos.

\section{Unsolved Puzzle --- the LSND Anomaly}
\label{sec_lsnd}

\begin{figure}[ht]
\centerline{\epsfig{width=0.7\textwidth, file=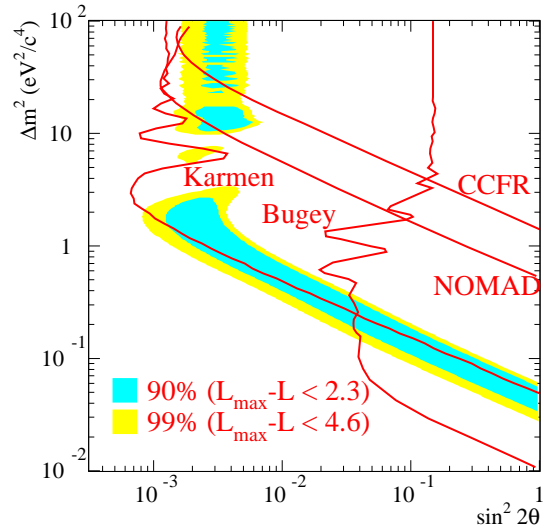}}
\caption{A $(\sin^22\theta,\Delta m^2)$
oscillation parameter fit
for the entire data sample,
$20<E_e<200$ MeV. The fit includes
primary $\bar \nu_\mu \rightarrow \bar \nu_e$
oscillations and secondary $\nu_\mu \rightarrow \nu_e$
oscillations, as well as all known neutrino backgrounds. The inner and
outer regions correspond to $90\%$ and $99\%$ CL allowed regions, while the
curves are $90\%$ CL limits from
the Bugey reactor experiment,
the CCFR experiment at Fermilab,
the NOMAD experiment at CERN, and
the KARMEN experiment at ISIS. From A.~Aguilar {\it et al.}  [LSND Collaboration],
Phys. Rev. D {\bf 64}, 112007 (2001). 
\label{fig_lsnd}}
\end{figure}

The LSND experiment\cite{lsnd} measured the neutrino flux produced by pion decay in flight ($\pi^+\to\mu^+\nu_{\mu}$) and antimuon decay at rest ($\mu^+\to e^+\nu_e\bar{\nu}_{\mu}$). It observed a small {\sl electron-type antineutrino} flux some 30 meters away from the production region.\cite{lsnd} The originally absent $\bar{\nu}_e$-flux can be interpret as evidence that $\bar{\nu}_{\mu}$ is transforming into $\bar{\nu}_{e}$ with $P_{\bar{\mu}\bar{e}}$ of the order a fraction of a percent. The data also contain a weak hint of a $\nu_e$ excess, which is both consistent with the $\bar{\nu}_{\mu}\to\bar{\nu}_e$ hypothesis and consistent with zero. If interpreted in terms of two-flavor neutrino oscillations, the LSND anomaly, combined with constraints imposed by several other experiments, points to a mass-squared difference $\Delta m^2_{\rm LSND}\sim 0.1-10$~eV$^2$, as depicted in Fig.~\ref{fig_lsnd}. 

The LSND result will be confirmed or refuted by the on-going MiniBooNE experiment, perhaps as early as the end of Summer 2005.\cite{miniboone} MiniBooNE consists of a standard $\nu_{\mu}$-beam experiment (from $\pi^+$ decay), with neutrino energies in the sub-GeV to GeV range and a baseline of around 500~m. Its beam (neutrino versus antineutrino) and systematics are quite distinct from the LSND setup. 

It is easy to understand why the LSND anomaly does not ``fit'' in the three flavor mixing scheme described in the previous section. With three neutrinos, one can define only two independent mass-squared differences, and these are completely determined by the solar, atmospheric, reactor, and accelerator data. As discussed earlier, both mass-squared differences are much smaller than $\Delta m^2_{\rm LSND}$. Given that the LSND results are yet to be confirmed by another experiment --- indeed, the Karmen 2 experiment,\cite{karmen} using a similar setup but with a shorter baseline ($L=18$~m), could have confirmed the LSND anomaly but did not observe any excesses, ruling out a significant portion of the LSND allowed parameter space (Fig.~\ref{fig_lsnd}) --- it is widely believe that  ``ordinary'' three-flavor oscillations are responsible for ``all-but-LSND-data,'' while the LSND anomaly could be due to more exotic new physics. Reinforcing this bias is the fact that,  if indeed present, the LSND anomaly requires a very small transition probability.

One possible solution to the LSND anomaly is to add extra, standard model gauge singlet neutrinos,\footnote{As discussed earlier, data from the LEP experiments,\cite{LEP_three} teach us that there are no extra very light ``neutrino'' degrees of freedom that couple to the $Z^0$-boson with standard model like couplings. Hence, additional light neutrinos are constrained to be gauge singlets, or sterile.} capable of mixing with the ordinary, or active, neutrinos. While this allows one to define at least three mass-squared differences, it is not guaranteed that one is capable of fitting all neutrino data with four (or more) neutrino mixing. Indeed, detailed analyses\cite{valle_new,strumia_sterile} suggest that four neutrino mixing schemes are either very poor or at best mediocre fits to all neutrino data. 

The reason for this can be qualitatively understood in the following way. There are two general neutrino mass-patterns capable of describing one large and two small mass-squared differences. These are referred to as the ``2+2'' and ``3+1'' scheme, and are depicted in Fig.~\ref{2+2,3+1}.

\begin{figure}[ht]
\centerline{\epsfig{width=0.6\textwidth, file=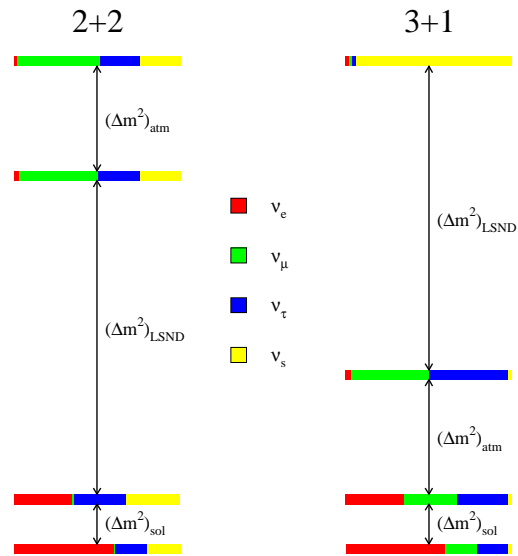}}
\caption{Two possible mass-patterns potentially capable of addressing all neutrino data, including those from LSND. The one on the left (right) is characteristic of a ``2+2'' (``3+1'') mass-scheme. \label{2+2,3+1}}
\end{figure}

The 2+2 schemes are disfavored for the following reason. Short baseline neutrino data constrain $|U_{\mu1}|,|U_{\mu2}|,|U_{e3}|,|U_{e4}|$ to be small.\footnote{Here, $|\Delta m^2_{12}|\equiv\Delta m^2_{\rm sol}$, $|\Delta m^2_{34}|\equiv\Delta m^2_{\rm atm}$.} If all of these are set to zero, atmospheric oscillations are driven by $|\nu_{\mu}\rangle \leftrightarrow \cos\zeta|\nu_{\tau}\rangle+\sin\zeta|\nu_{s}\rangle$ mixing, where $\nu_s$ is a sterile neutrino and $\zeta$ is a mixing angle that characterizes the sterile component of the atmospheric neutrino ``pair'' of the muon-type neutrino. By unitarity, solar oscillations are driven by $|\nu_{e}\rangle\leftrightarrow-\sin\zeta|\nu_{\tau}\rangle+\cos\zeta|\nu_s\rangle$ mixing. 

Both solar and atmospheric data constrain $\zeta$. Atmospheric neutrino data are sensitive to a sterile component via Earth matter effects, and due to the fact that tau-type neutrinos also interact with the detector. Solar data are sensitive to a sterile component via matter effects in the Sun and in the Earth, and also due to the fact that tau-type neutrinos interact via neutral current interactions inside of SNO and Super-Kamiokande. Both solar and atmospheric data see no evidence for a sterile component. Hence, atmospheric data set an upper bound for $\sin^2\zeta$, while the solar data require a small $\cos^2\zeta$. By combining both data sets, $\cos^2\zeta+\sin^2\zeta$ is already constrained to be less than one,\cite{valle_new,strumia_sterile} ``ruling out'' the 2+2 scheme. Some attempts have been made at understanding what happens when the conditions $|U_{\mu1}|=|U_{\mu2}|=|U_{e3}|=|U_{e4}|=0$ are lifted consistent with experimental bounds (see, for example, Ref.~\refcite{weiler_etal}) but one expects that these sub-leading effects will not lead to a significantly better fit.

The 3+1 schemes fit the atmospheric and solar data just fine, given that sterile neutrino effects are just a small perturbation to the ordinary three neutrino fit to all-but-LSND data. They run into some trouble when it comes to short-baseline searches for $\nu_e$ and $\nu_{\mu}$ disappearance driven by the LSND frequency (see Fig.~\ref{fig_lsnd}). LSND $\nu_{\mu}\leftrightarrow\nu_e$ oscillations are given by\footnote{Here, $\Delta m^2_{\rm LSND}\simeq|\Delta m^2_{i4}|,~\forall i=1,2,3$.} 
\begin{equation}
P_{e\mu}\simeq 4|U_{e4}|^2|U_{e4}|^2\sin^2\left(\frac{\Delta m^2_{\rm LSND } L}{4E}\right),
\end{equation}
while the survival probability of a species $\alpha=e,\mu$, for ``short'' $L$, is given by
\begin{equation}P_{\alpha\alpha}\simeq 1-4|U_{\alpha4}|^2(1-|U_{\alpha4}|^2)\sin^2\left(\frac{\Delta m^2_{\rm LSND} L}{4E}\right).
\end{equation}
The absence of electron-type and muon-type neutrino disappearance constrains $|U_{e4}|^2,|U_{\mu4}|^2$ to be small, while the LSND data require $|U_{e4}|^2\times|U_{\mu4}|^2$ to be larger than a fraction of a percent.\cite{31_trouble} The tension in the current data is not enough to rule out the 3+1 schemes, but it does lead to a mediocre fit.\cite{valle_new,strumia_sterile}

Five neutrino mixing schemes have also been explored (see, for example, Ref.~\refcite{five}). These look like ``3+1+1'' schemes (``2+2+1'' schemes do not fare much better than 2+2 schemes) and are designed in a such a way that the tension between the short-baseline and the LSND data is alleviated. With five neutrinos, it is possible to fit all the neutrino data properly, but some worry that the choices for mixing parameters and mass-squared differences are rather ``finely tuned.''

More exotic solutions to the LSND anomaly have been proposed, and none of them seem to fit all data particularly well. Here I list some of them.

The possibility that there are rare lepton-flavor violating $\mu^+\to e^+\nu_{\alpha}\bar{\nu}_e$ decays\cite{babu_pakvasa} could explain the LSND data as long as the branching ratio for the flavor-violating decay was of order a fraction of a percent. Such decays, however, should also have been observed by the Karmen experiment, which disfavored this hypothesis at around the 90\% confidence level.\cite{karmen_mu} Recently available precision data of the Michel electron energy spectrum\cite{TWIST} seem to safely rule out flavor changing muon decays as a solution to the LSND anomaly. 

Postulating that neutrinos and antineutrinos have different masses and mixing angles\cite{cptv} received a significant amount of attention in the past three years. The original idea was inspired by the fact that solar data required the disappearance of electron-type {\sl neutrinos}, while those from LSND required the appearance of electron-type {\sl antineutrinos}. If neutrinos and antineutrinos oscillated at different frequencies (different $\Delta m^2$), all data could be rendered compatible. Aside from all sorts of theoretical issues, the original CPT-violating setup was ruled out when KamLAND published the first evidence for antineutrino oscillations at solar frequencies. A second manifestation of CPT-violating solutions the the LSND data consisted of postulating that atmospheric oscillations in the antineutrino sector were driven by $\Delta m^2_{\rm LSND}$. This possibility is strongly disfavored (at the three sigma level) by the atmospheric data.\cite{no_cptv}

Given the fact that none of the solutions to the LSND anomaly proposed to date seem completely satisfactory, it is fair to say that if MiniBooNE confirms the observations made by LSND, there is a good chance we have uncovered a novel physical phenomenon, {\it i.e.}, we are yet to figure out what the LSND result is teaching us. This being the case, if the LSND anomaly is confirmed, all the necessary experimental and theoretical efforts will most likely concentrate, first, on uncovering the mechanism responsible for the LSND flavor change. This will likely require (i) detailed analysis of all available data, (ii) new, compelling ideas, and (iii) a series of other experimental neutrino efforts, capable of mapping out the LSND/MiniBooNE potential parameter spaces. We are still not sure what these should be!

\section{What Have We Learned?}
\label{theend}

If one were to summarize in one sentence what we have learned after six years of remarkable experimental results, this would be it: 
\begin{center}
Neutrinos have (very, very tiny) masses.
\end{center}
Massive neutrinos constitute the only palpable evidence we have that the standard model of electroweak, and strong interactions (SM) does not describe all strong, electromagnetic and weak phenomena. 

\begin{figure}[th]
\centerline{\psfig{file=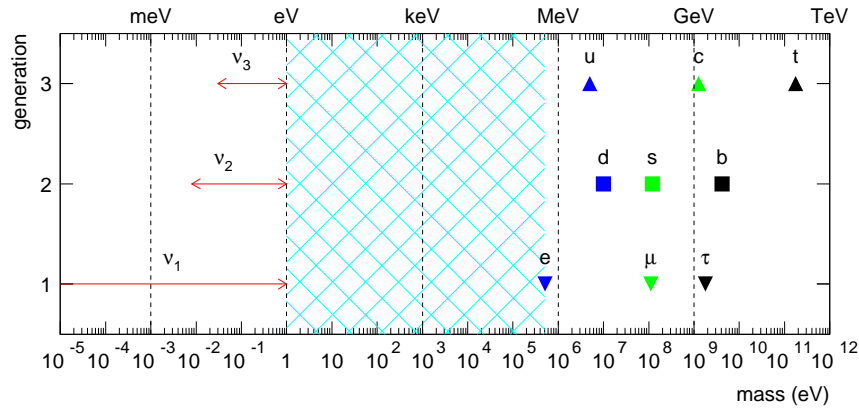,width=1\textwidth}}
\vspace*{8pt}
\caption{Masses of all known fundamental fermions. A normal mass-hierarchy has been assumed --- $m_{\nu_1}^2<m_{\nu_2}^2<m_{\nu_3}^2$ ---  together with a rather conservative upper bound $m_{\nu_i}^2<1$~eV $\forall i=1,2,3$. The light, hatched region indicates the six-orders-of-magnitude ``desert'' between the largest possible neutrino mass and the electron mass.}
\label{allmasses}
\end{figure}

Fig.~\ref{allmasses} depicts the value of the masses of all known fundamental fermions. Note the logarithmic scale. Fermion masses are very hierarchical --- it takes over thirteen orders of magnitude to fit them all in one plot! Quark masses span  five orders of magnitude, while charged fermion masses span over three orders of magnitude. We don't know why fermion masses are distributed in this way. 

It is remarkable that the ratio of the largest possible neutrino mass to the lightest known charged fermion mass (the electron mass) is at least one order of magnitude less than the ratio of the electron mass to the top quark mass. Furthermore, while the electron--top ``gap'' is populated by all other charged leptons and quarks, the heaviest-neutrino--electron ``gap'' seems to be deserted.  We also don't know why this is the case, but it does seem to be Nature's way of saying that there is something special about neutrino masses. It may be that neutrino masses are qualitatively different from charged fermion masses.

Before proceeding, I'll very briefly summarize the SM, and what I mean, here, by going beyond it (see also Ref.~\refcite{willen}). The SM is a Lorentz invariant quantum field theory, and its renormalizable Lagrangian is uniquely determined once one specifies its internal symmetries (gauged $SU(3)_c\times SU(2)_L\times U(1)_Y$ invariance) and particle content ($Q,u,d,L,e$, the matter fields, plus $H$, the Higgs doublet scalar field). The fact that the Lagrangian is renormalizable implies that, naively, the SM is valid up to arbitrarily high energy scales (ignoring gravity, etc). It is easy to check that given the SM as defined above, neutrinos are strictly massless. 

There are several ways of modifying the SM and allowing nonzero neutrino masses. The amount of experimental information available is, however, still insufficient to allow a particular candidate ``new SM'' to be chosen over another, but there is reason to believe that more information is on the way in the near (few years) to intermediate (several years) future. Here, I'll very briefly present two ``minimal paradigms.''

Arguably, the simplest way to ``add'' neutrino masses to the SM is to give up on the renormalizability of the Lagrangian. This allows one to add (an infinite number of) irrelevant operators consistent with the symmetries:
\begin{equation}
{\rm L}_{\rm new}={\rm L}_{\rm SM}-\lambda^{\alpha\beta}\frac{L_{\alpha}HL_{\beta}H}{2M}+ O\left(\frac{1}{M^2}\right).
\label{eq:majorana}
\end{equation}
All fermion fields are understood to be Weyl fermions, such that, for example, the charged lepton Yukawa operator is written as $LHe$, where $e$ is the (positively) charged-lepton $SU(2)_L$ singlet field. 

Two facts are remarkable. One is that $(LH)^2$ is the {\sl only} type of dimension-five operator allowed by the SM gauge invariance and particle content.\cite{weinberg} The other is that, as long as $M$ is much larger than $\langle H\rangle$, the Higgs 
vacuum expectation value, the only observable consequence of Eq.~(\ref{eq:majorana})\footnote{This is not necessarily correct. One also has to worry about dimension six operators that lead to baryon number violation and a finite lifetime for the proton.} is that neutrinos get a nonzero  mass after electroweak symmetry breaking: $m_{\nu}=\lambda\langle H\rangle^2/M$. An extra ``bonus'' is that neutrino masses are naturally much smaller than all other fermion masses $m_f\propto\langle H\rangle$ by a factor $\langle H\rangle/M$.

Another important consequence of Eq.~(\ref{eq:majorana}) is that lepton number is not a good symmetry ($(LH)^2$ breaks lepton number by two units). Lepton number violation is ``encoded'' in the fact that the neutrinos are Majorana fermions. 

$M$ can be described, roughly, as the energy scale above which  Eq.~(\ref{eq:majorana}) is no longer valid. There is very little information regarding the magnitude of $M$, but one can set an upper bound for $M$ by assuming that $\lambda\sim 4\pi$,\cite{willen2} {\it i.e.}, by assuming that the physics replacing Eq.~(\ref{eq:majorana}) at the scale $M$ is strongly coupled. In this case 
\begin{equation}
M\lesssim4\pi\frac{\langle H\rangle^2}{m_{\nu}}\sim10^{15}~\rm GeV\left(\frac{100~\rm meV}{m_{\nu}}\right).
\end{equation}
If Eq.~(\ref{eq:majorana}) is indeed the correct low-energy description of Nature, neutrino masses represent the first direct evidence that the SM is an effective field theory, valid up to an energy scale less that $10^{15}$~GeV (or so), which is, in turn, much less than the Planck scale.\cite{willen2,degouvea_valle} It is also impressive that the upper bound above coincides, qualitatively, with the energy scale where all three running gauge coupling constants of the SM seem to meet,  $M_{\rm GUT}\sim 10^{15-16}$~GeV.

There are several different proposals for the physics that replaces Eq.~(\ref{eq:majorana}). The most famous one is the seesaw mechanism,\cite{seesaw} described in detail in Ref.~\refcite{willen}.

A completely different option is to assume that neutrinos are, similar to all charged matter fields, Dirac fermions. In this case, in order to render the neutrinos massive, it suffices to add extra SM gauge singlet Weyl fermions $N_i$ (``right-handed neutrinos''), and Yukawa couplings between $H$, $L_{\alpha}$, $N_i$:
\begin{equation}
{\rm L}_{\rm new}={\rm L}_{\rm SM}-y_{\alpha i}L_{\alpha}HN_i+H.c.~.
\label{eq:dirac}
\end{equation} 
After electroweak symmetry breaking, the neutrino mass matrix is given by $m_{\nu}=y\langle H\rangle$, similar to the up-type and down-type quark mass matrices and the charged lepton mass matrix. The magnitude of the neutrino masses requires $y\lesssim 10^{-12}$, at least six orders of magnitude smaller than the electron Yukawa coupling. It is clear that a natural explanation for the smallness of the neutrino mass is {\sl not} contained in Eq.~(\ref{eq:dirac}). On the positive side, Eq.~(\ref{eq:dirac}) is renormalizable, meaning that this new SM version is, naively, valid up to arbitrarily high energy scales. 

Modulo a natural explanation for the size of the neutrino mass, one could try to argue that Eq.~(\ref{eq:dirac}) is a rather innocuous addition to the SM Lagrangian. I believe this is not the case. Eq.~(\ref{eq:dirac}) is not the most general, renormalizable Lagrangian consistent with the symmetries of the SM. Once the fields $N_i$ are introduced, the dimension-three Majorana mass operators $\frac{1}{2}M_N^{ij}N_iN_j$ should also have been introduced, given the fact that $N_i$'s are gauge singlets. One needs, therefore, to modify the {\sl symmetry structure} of the SM in order to forbid a Majorana mass term for the right-handed neutrinos. One simple way of doing this is to add to the SM an internal global symmetry, {\it e.g.} $U(1)_{B-L}$, where $B$ stands for baryon number, and $L$ for lepton number.\footnote{$U(1)_{B-L}$ is not anomalous, meaning it is not broken by non-perturbative quantum mechanical effects, unlike $U(1)_B$ or $U(1)_L$.} Note that, in the massless-neutrino SM, $U(1)_{B-L}$ is an {\sl accidental} global symmetry, {\it i.e.}, it arises as a consequence of the imposed gauge symmetries and the particle content. Among other things, this means that there was, {\it a priori}, no reason to believe that it needed to be conserved by allowed SM extensions, including SUSY, quantum gravitational effects, etc. If Eq.~(\ref{eq:dirac}) is indeed the correct description of neutrino masses, $U(1)_{B-L}$ needs to be ``upgraded'' to an imposed fundamental global symmetry, and it is expected to be preserved by, say, quantum gravity, etc.

Several distinct mechanisms for explaining naturally light Dirac neutrino masses exist, for example, in the various models with either ``large'' or ``warped'' extra dimensions.\cite{add_nus,rs_nus,fat_nus} This is rather convenient, given that most realizations of these models have a rather low ultraviolet cutoff, meaning that these models are effective field theories that need to be replaced by unknown new physics at energy scales well below the Planck mass (more often, close to the electroweak breaking scale). Therefore, in order to avoid generic dimension-five operators as in Eq.~(\ref{eq:majorana}) suppressed by $M\gtrsim 1$~TeV and hence unacceptably large Majorana neutrino masses, one is ``required'' to impose $U(1)_{B-L}$ (or something similar) as a fundamental global symmetry.\footnote{In all fairness, this is not a unique ``feature'' of extra-dimensional theories. The MSSM, for example, runs into similar problems if R-parity is not imposed as a fundamental symmetry. This is, of course, a consequence of the fact that $U(1)_{B-L}$ is an {\sl accidental symmetry} of the SM.}

\subsection{The Faith of Lepton Number}

Except for the nature of the neutrinos, both neutrino mass paradigms --- Eq.~(\ref{eq:majorana}) and Eq.~(\ref{eq:dirac}) --- predict one and the same thing: neutrinos have mass. In order to make significant progress in understanding neutrino masses, it is clear that we need to establish, by some means, whether neutrinos are Dirac or Majorana fermions. 

The way to test whether neutrinos are Majorana fermions is to look for processes that violate lepton number, including forbidden decays like $K^+\to\pi^-\mu^+\mu^+$ and/or scattering processes, like $\nu_e+p\to e^++n$. It turns out that if the neutrinos are Majorana fermions, all such processes are expected to occur with nonzero probability. The reason for this is simple. If neutrinos are Majorana fermions, $U(1)_L$ (and $U(1)_{B-L}$) are not good symmetries of the SM, and nothing would prevent, say, $K^+\to\pi^-\mu^+\mu^+$ from happening. The converse is also true --- if there are physical processes that violate baryon number minus lepton number, then neutrinos are Majorana fermions.\cite{0nubb_majorana}

If, however,  the neutrino masses are the only source of lepton number violation, lepton number violating effects are, numerically, tiny. This is easy to see. If the neutrino mass $m_{\nu}$ is the only parameter that ``knows'' about lepton number violation, the amplitude for {\sl any} lepton-number violating process should vanish when $m_{\nu}\to0$ (when the accidental $U(1)_{B-L}$ symmetry is restored) such that $A_{L\hspace{-1.6mm}\slash}\propto\left(\frac{m_\nu}{E}\right)^{n}$, where $n$ is positive and $E$ is the typical energy scale involved in the process of interest. Because neutrino masses are tiny compared to any reasonable value of $E$, the rate of lepton number violating processes is expected to be hopelessly small.

Given the current bounds on neutrino masses and mixing, the ``only hope'' for probing lepton number violation mediated by Majorana neutrino masses with enough sensitivity is to look for neutrinoless double-beta decay, $0\nu\beta\beta$.\cite{0nubb}

\begin{figure}[ht]
\centerline{\epsfig{width=1\textwidth, file=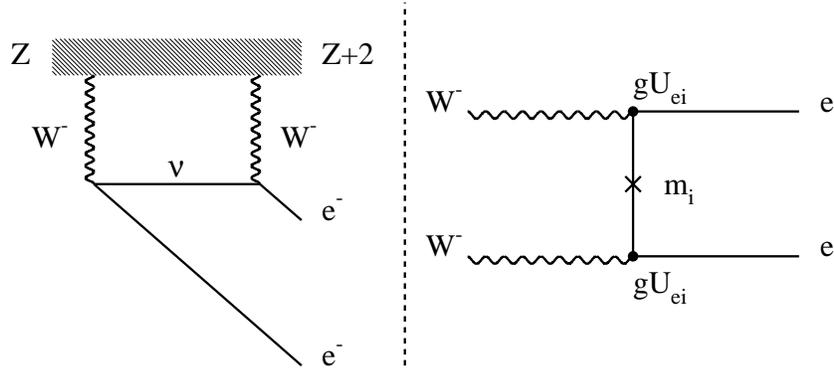}}
\caption{LEFT -- Diagram contributing to $0\nu\beta\beta$. The hatched region indicates the ``nuclear physics'' part of the process. RIGHT -- The high energy ``core'' of $0\nu\beta\beta$, $W^-+W^-\to e^-+e^-$ via $\nu_i$ exchange ($i=1,2,3$, $m_{\nu_i}\equiv m_i$). Here, $g$ is the weak coupling, $U$ is the lepton mixing matrix, and the cross indicates a fermion mass insertion. \label{fig:0nubb}}
\end{figure}

Two-neutrino double beta decay ($2\nu\beta\beta$) is a nuclear process through which a $Z$-charged nucleus decays ``directly'' to a $Z+2$-charged nucleus:
\begin{equation}
Z\to(Z+2)+e^-+e^-+\bar{\nu}_e+\bar{\nu}_e.
\end{equation}
Such processes have been observed for several nuclei, including $^{76}$Ge, $^{100}$Mo, $^{130}$Te, etc. Typical half-lives are well above $10^{18}$ years (one hundred million times the age of the Universe!). Similarly, $0\nu\beta\beta$ is characterized by
\begin{equation}
Z\to(Z+2)+e^-+e^-,
\end{equation}
and violates lepton number by two units. It can be interpreted as a $2\nu\beta\beta$ process where the two antineutrinos ``annihilate'' into vacuum. The diagram that describes neutrino-mass-induced $0\nu\beta\beta$ is depicted in Fig.~\ref{fig:0nubb}(LEFT). The high energy physics ``core'' of the process is depicted in Fig.~\ref{fig:0nubb}(RIGHT), and consists of the lepton-number violating scattering process $W^-+W^-\to e^-+e^-$.

The amplitude for $0\nu\beta\beta$ is proportional to 
\begin{equation}
A_{0\nu\beta\beta}(E)\propto\sum_i U_{ei}^2\frac{m_i}{E}\equiv\frac{m_{\beta\beta}}{E},
\end{equation}
where $m_{\beta\beta}$ (also known as $m_{ee}$) is referred to as the effective neutrino mass for $0\nu\beta\beta$ and $E$ is some fixed energy for the process. As advertised, $A_{0\nu\beta\beta}$ is directly proportional to the neutrino mass.

It is important to note that $m_{\beta\beta}$ is a complex quantity, and that it depends on specific combinations of the Dirac and Majorana phases $\delta,\xi,\eta$, defined in Sec.~\ref{solve_puzzle}. Furthermore, $m_{ee}$ could end up being much smaller than the typical $m_i$ if its different components add up destructively. The value of $m_{ee}$ depends not only on the neutrino mass-squared differences and mixing parameters, but also on the magnitude of the masses. This is why failed searches for neutrinoless double beta decay set bounds on the scale of the neutrino mass.

Assuming, say, an inverted mass-hierarchy
\begin{eqnarray}
\left|m_{\beta\beta}\right|&=&\cos^2\theta_{13}\left|\cos^2\theta_{12}m_1+\sin^2\theta_{12}m_2e^{i\phi}\right|, \\
&\simeq& \cos^2\theta_{13}m_1\left|\cos^2\theta_{12}+\sin^2\theta_{12}e^{i\phi}\right|, \\
&\ge&\cos^2\theta_{13}\left|\cos2\theta_{12}\right|\sqrt{\Delta m^2_{13}},
\end{eqnarray}
where $\phi$ is the relative phase between $U_{e1}^2$ and $U_{e2}^2$. Note that, because we know that the solar mixing angle is not maximal, in the case of an inverted mass hierarchy we can set a lower bound for $m_{ee}\gtrsim 0.02$~eV. 

In the case of a normal mass-hierarchy (and assuming, say $|m_1|\ll|m_2|$), $m_{ee}\gtrsim \left|\cos2\theta_{12}\right|\sqrt{\Delta m^2_{12}}\sim 0.003$ for ``small'' $\theta_{13}$, while large cancellations can occur if the values of $\theta_{13}$, $m_1$, and the relative CP-odd phases are ``just-right.''  

The experimental search strategy is clear. One should gather a high-statistics sample of double beta decay candidates, and study the end-point of the energy spectrum of the two-electron final state. $0\nu\beta\beta$ events will accumulate at the edge of the allowed phase space for $2\nu\beta\beta$. Current experiments bound $m_{ee}<\rm several~\times 10^{-1}$~eV (with large uncertainties from the nuclear matrix elements).\cite{PDG} As of a few years ago, a controversial reanalysis of the Heidelberg-Moscow $^{76}$Ge data uncovered a positive hint for $0\nu\beta\beta$.\cite{Klapdor} This evidence would correspond to $m_{ee}\in[0.2,0.6]$~eV at the 99\% confidence level.\cite{Klapdor} 

Next-generation $0\nu\beta\beta$ searches have been proposed, and aim to be sensitive to $m_{ee}>0.1$~eV (and capable of verifying the claims of Ref.~\refcite{Klapdor}). One can expect some of these proposals to be fully funded and running in the next few years.\cite{APS} Upgrades to these proposals claim to able to reach $m_{ee}>0.01$~eV. It is a good bet that we will either discover $0\nu\beta\beta$ or rule out Majorana neutrinos with an inverted mass hierarchy in a little over ten years.\footnote{As I tried to illustrate throughout the previous sections, progress in neutrino physics is typically quite slow, but more than often worth the wait!}

\subsection{Brief Concluding Remarks}

The discovery of neutrino masses has not only provided the first evidence of new physics, but also opened up the door for several new theoretical and experimental developments. I only managed to mention a tiny fraction of those here, and completely neglected to comment on very important current areas of research, including neutrinos in astrophysics and cosmology, and the neutrino mixing puzzle (why is lepton mixing so different from quark mixing?).

We have only just begun to decipher what  Nature is saying through the ghost-like neutrinos, and one can count on a very bright (and not too far away!) future full of surprises and, hopefully, a deeper and more satisfying understanding of fundamental physics. 

\section*{Acknowledgments}
I am happy to thank John Terning, Carlos Wagner, and Dieter Zeppenfeld, the organizers of the 2004 ``Physics in $D>4$'' TASI, for the invitation to present these lectures and for putting together a nice program, and am grateful to K.~T. Mahanthappa and Tom DeGrand for their hospitality. I also thank the TASI participants and the students at Northwestern University for many questions and comments, and Scott Willenbrock for nice discussions on the fundamentals of neutrino physics. 

\appendix

\section{Homework Problems}

These are some homework problems I assigned during a Spring 2004 course on neutrino physics at Northwestern University. For solutions and more details, see http://lotus.phys.nwu.edu/$\sim$degouvea/neutrinos.html . Some of the problems included here were taken from half of a one semester course ministered by Hitoshi Murayama in the Fall of 1998 at UC Berkeley. Enjoy!

\renewcommand{\theequation}{A.\arabic{equation}}

\begin{enumerate}

\item The charged pion decays almost 100\% of the time into a muon and a (muon-type) neutrino ($\pi^{+}\to\mu^+\nu_{\mu}$). In the reference frame where the parent pion is at rest, compute the muon energy as a function of the muon-mass ($m_{\mu}$), the charged pion mass ($m_{\pi}$), and the neutrino mass ($m_{\nu}$). What is the absolute value of the muon momentum (tri)vector? Numerically, what is the relative change of the muon momentum between $m_{\nu}=0$ and $m_{\nu}=0.1$~MeV? It is remarkable that the muon momentum from pion decay at rest has been measured at the $3.4\times 10^{-6}$~level (Phys. Rev. D{\bf 53}, 6065 (1996)). This provides the most stringent current constraint on the ``muon-neutrino mass." We will later discuss the meaning of this bound. 

\item At small enough energies ($\sqrt{s}<O(100)$~GeV), the neutrino cross section is approximately given by $\sigma_{\nu}\sim G_F^2s/\pi$, where $G_F$ is the Fermi constant and $s=(p+P)^2$ is the square of the center-of-mass energy of the neutrino (with four-momentum $p_{\mu}$) plus target (with four-momentum $P_{\mu}$) system. 

(a) Estimate the cross section, in cm$^2$ for neutrino--electron scattering and neutrino--neutron scattering when $e$ and $n$ are at rest and the neutrino energy $E_{\nu}=10$~MeV.

(b) Estimate the mean free path  of a 10~MeV neutrino through lead, in A.U. [1~A.U.  (one Astronomical Unit), is the average Earth--Sun distance, equal to $1.5\times 10^{11}$~m (or 500 light-seconds).]


\item To understand the effect of neutrino oscillations (consider two flavor $\nu_{\mu}\leftrightarrow\nu_{\tau}$ transitions) on the atmospheric muon-neutrino data, numerically calculate and draw histograms of the average muon neutrino survival probability in ten equal-size bins of $\cos\theta_z$, where $\theta_z$ is the angle between the neutrino direction and the vertical-axis at the detector's location ($\theta_z=0$ for neutrinos coming straight from above, and $\theta_z=\pi$ for neutrinos coming from below). Make one histogram for $E_{\nu}=0.2$~GeV, 2~GeV, and 20~GeV plus $\Delta m^2=2.5\times 10^{-4}$~eV$^2$, $2.5\times 10^{-3}$~eV$^2$, and $2.5\times 10^{-2}$~eV$^2$, for a grand total of nine plots. Assume throughout that the mixing is maximal, i.e., $\sin^2 2\theta=1$, and that neutrinos are produced 20~km above the surface of the Earth.

\item Read the article "Super-Kamiokande Atmospheric Neutrino Results" by T.~Toshito, hep-ex/0105023. It contains an almost up-to-date summary of the atmospheric neutrino data (not much more data has been collected since, for reasons that I'll mention briefly in class). A talk by T. Kajita, presented at the Neutrino 1998 Conference, may also prove helpful in understanding some of the Super-Kamiokande terminology: hep-ex/981001.

(a) From Table 1, compute the value of the ``ratio-of-ratios" $R$ (the measured $\nu_{\mu}$ to $\nu_e$ flux ratio divided by the theoretical calculation) for sub-GeV and multi-GeV single ring events, and compare them to the numbers quoted in the paper. How do these numbers compare to 1, to each other, and to the ratio of observed partially contained events to the Monte Carlo calculation (this are all muon-type events, and consist of events whose average energy is larger than that of the multi-GeV events)? Discuss possible interpretations for these discrepancies.

(b) Look at Figure 1, and compare with the results you got in problem 1. Can you verify that $\Delta m^2\sim2.5\times 10^{-3}$~eV$^2$ and $\sin^22\theta\sim1$ is a good fit to the data (200~MeV is characteristic of sub-GeV events, 2~GeV is typical of multi-GeV events, and 20~GeV is typical of upward stopping muons. The fourth category, upward-through-going muons, has an average energy above 100~GeV)? In particular, explain why there is almost no depletion for $\cos\theta_z>0.2$ in the multi-GeV data, but some depletion in the sub-GeV data.  

(c) Use the number of observed sub-GeV ``$e$-like" events (as these seem to agree well with Monte Carlo predictions) to obtain an order of magnitude estimate of the electron neutrino flux (neutrinos per unit time and unit area). The cross section for detecting neutrinos at this energy range is roughly 5~fb.  

\item {\bf Understanding SNO data} --- Read Phys. Rev. Lett. {\bf 89}, 011301 (2002), which describes the results of the SNO (Sudbury Neutrino Observatory) experiment [nucl-ex/0204008]. In page 4, the collaboration quotes the measured values of the ``solar neutrino flux,'' obtained by using different physical processes: $\phi_{\rm CC}$ is determined from the Charged Current reaction $\nu+d\to p+p+e^-$ ($d$ is a deuteron nucleus), $\phi_{\rm NC}$ is determined from the Neutral Current reaction $\nu+d\to n+p+\nu$, while $\phi_{\rm ES}$ is determined from the Elastic Scattering reaction $\nu+e^-\to \nu+e^-$. These flux-measurements are obtained assuming that only electron-type neutrinos are coming from the Sun.

The key point is that the Charged Current process is only sensitive to electron-type neutrinos, the Neutral Current reaction is flavor blind (i.e. does not care whether the neutrino is of the electron-, muon- or tau-type), while the Elastic Scattering reaction is sensitive  to electron-type neutrinos and muon/tau-type neutrinos in a different way. At the energy range of interest to the SNO experiment, the ratio of the elastic scattering cross sections is $\sigma_{\rm ES}(\nu_{a}+e)/\sigma_{\rm ES}(\nu_e+e)=0.154$, and very close to being energy independent. Here, $a=\mu$~and/or~$\tau$. [At solar neutrino energies, there is no way of distinguishing muon-type from tau-type neutrinos. Here, you can simply refer to them as $\nu_a$, where $a$ stands for `active.']

(a) Rewrite $\phi_{\rm CC}$, $\phi_{\rm NC}$ and $\phi_{\rm ES}$ in terms of $\phi_e$ and $\phi_{a}$, the flux of electro-type solar neutrinos and the flux of muon/tau-type solar neutrinos. Given the experimental results obtained by SNO, compute $\phi_e$ and $\phi_{a}$. [This system is overconstrained --- there are three equations and two unknowns. You can either make sure that the three measurements are consistent (reproducing something like Fig.3 in the paper is a good idea!), or you can perform a quick fit to the three measurements. If you choose to do this, for simplicity, add the statistical and systematic errors in quadrature, and assume that this combined error is Gaussian. I recommend the second option (it is important to learn how to combine data and extract the value of physical parameters. Please let me know if you have no idea what I am talking about!).] Compare your results with those obtained by the collaboration, quoted in page 5. The fact that $\phi_{a}\neq 0$ is, currently, the most concrete evidence we have of neutrino flavor conversion, since there are no physical processes capable of producing non-electron-type neutrinos inside the Sun!

(b) Assume that the survival probability of electron-type neutrinos $P_{ee}$ is energy dependent, so you can rewrite $\phi_e=P_{ee}\phi_{\odot}$, $\phi_{a}=(1-P_{ee})\phi_{\odot}$, where $\phi_{\odot}$ is the total neutrino flux from the Sun. From the SNO data, calculate the values of $P_{ee}$ and $\phi_{\odot}$.
Note that $P_{ee}<0.5$ is indicative of ``strong" matter effects inside the Sun combined with the fact that the electron-type neutrino is predominantly light, i.e., $\sin^2\theta<0.5$.

\item {\bf Day-Night Effect} --- Solar neutrino oscillations can also be modified by the fact that, during the night, the neutrinos have to cross some significant amount of the Earth in order to reach the detectors. Hence, the oscillation probability is different for neutrinos arriving during the day and the night (experiments with real-time event reconstruction capabilities search for a day-night asymmetry in the measured solar neutrino flux).

To understand this effect, assume that solar neutrinos arrive at the surface of the Earth in the  $|\nu_2\rangle$ state (a mass eigenstate). This is true of $^8$B solar neutrinos as long as few$\times10^{-9}~{\rm eV^2} < \Delta m^2 < {\rm few}\times10^{-5}~{\rm eV}^2$ and $\sin^2\theta$ is not too small ($\sin^2\theta>0.1$ is safe).

\item It is easy to show that the $\nu_{\mu}\to\nu_e$  oscillation probability for three active flavors in vacuum can be written as
\begin{equation}
P(\nu_{\mu}\to\nu_e)=\sum_{i,j=1}^3U^*_{ei}U_{\mu i}U_{ej}U_{\mu j}^*\exp\left(-i\frac{(m^2_i-m^2_j)L}{2E}\right).
\end{equation}
This form proves useful to address the following questions:

(a) Show that time-reversal invariance is not necessarily conserved, {\it i.e.}, that $P(\nu_{\mu}\to\nu_e)\neq P(\nu_e\to\nu_{\mu})$ unless $U$ is a real matrix. Can you find a simple expression for $P(\nu_{\mu}\to\nu_e) - P(\nu_e\to\nu_{\mu})$?

(b) The mixing matrix for antineutrinos is the same as the one for neutrinos, except for $U\leftrightarrow U^*$. Show that CP-invariance is not necessarily conserved, {\it i.e.}, that $P(\nu_{\mu}\to\nu_e)\neq P(\bar{\nu}_{\mu}\to\bar{\nu}_e)$.  Can you find a simple expression for 
$P(\nu_{\mu}\to\nu_e)-P(\bar{\nu}_{\mu}\to\bar{\nu}_e)$? How does it relate to the one you may have obtained in (a)?

(c) Show that CPT invariance is conserved, {\it i.e.}, $P(\nu_{\mu}\to\nu_e)=P(\bar{\nu}_e\to\bar{\nu}_{\mu})$.

\item {\bf SN1987A} --- In February 1987, neutrinos from a Supernova that exploded in the Large Magellanic Cloud, located 50~kpc away from the Earth, reached the Kamiokande and the IMB experiments. This neutrino burst was reported in K.~Hirata {\it et al.}, Phys. Rev. Lett. {\bf 58}, 1490 (1987) and R.M.~Bionda {\it et al.}, Phys. Rev. Lett. {\bf 58}, 1494 (1987). Read the two papers and address the following questions:

(a) Both experiments detect electron antineutrinos via $\bar{\nu}_e p\to e^+n$. Calculate the incoming antineutrino energy, in the reference frame where the target protons are at rest, as a function of the recoil angle and energy of the positron, and the neutron and proton masses. You may set the positron mass to zero. Using the table of events in the Kamiokande paper, compute the energies of the antineutrinos that were deteced by the Kamiokande experiment. What is the highest (lowest) observed antineutrino energy?

(b) Use the result from (a) to obtain an upper bound on the ``electron antineutrino mass" (don't worry about mixing). The reasoning is the following: if neutrinos have mass, neutrinos with different energies propagate with slightly different velocities. Hence, the higher energy neutrinos should have arrived at the detectors before the lower energy ones. Compute the relative arrival time of two antineutrinos with two different energies (assume that the neutrino mass is a lot smaller than any neutrino energy in the problem). The time distribution of the events observed by Kamiokande is consistent with the expected antineutrino energy spread (a few seconds). From this fact, place an upper bound on the neutrino mass.
[The last three events arrives more than nine seconds after the first event, and there is still a debate regarding whether these are really from the Supernova. Discard them, and consider only the first nine events.]

Useful information: 1 parsec is about $3\times 10^{16}$~m.

\end{enumerate}

%
%
%
%

\end{document}